\documentclass[a4paper,useAMS,usenatbib]{mn2e}

\usepackage{graphicx}	
\usepackage{amsmath}	
\usepackage{amssymb}	
\usepackage{subfigure}
\usepackage{mathrsfs}

\title[Implications of Crab Inner Knot]{On the Implications of Recent Observations of the Inner Knot in the Crab Nebula}

\author[Y. Yuan \& R. Blandford]{
Yajie Yuan,$^{1}$\thanks{E-mail: yuanyj@stanford.edu (YY)}
Roger Blandford$^{1}$\thanks{E-mail: rdb3@stanford.edu (RB)}
\\
$^{1}$Kavli Institute for Particle Astrophysics and Cosmology, SLAC and Stanford University, Stanford, CA 94305, USA
}

\begin{document}

\date{Accepted XXX. Received YYY; in original form ZZZ}

\pagerange{\pageref{firstpage}--\pageref{lastpage}} \pubyear{2015}

\maketitle

\label{firstpage}

\begin{abstract}
Recent observations of the Crab Nebula \citep{Rudy:2015aa} have maintained its reputation for high energy astrophysical enlightenment and its use as a testbed for theories of the behaviour of magnetized, relativistic plasma. In particular, new observations of the inner knot located $0.65\arcsec$ SE from the pulsar confirm that it is compact, elongated transversely to the symmetry axis and curved concave towards the pulsar. 60 percent polarization has been measured along the symmetry axis \citep{Moran:2013aa}. The knot does not appear to be involved in the gamma ray flares. The new observations both reinforce the interpretation of the knot as dissipation of the pulsar wind at a strong shock and challenge the details of existing models of this process. In particular, it is argued that the compactness, high polarization and curvature are difficult to reconcile with simple relativistic shock models. Alternative possibilities include deflection of the outflow ahead of the shock and spatial variation in which the knot is interpreted as a caustic. Some future observations are proposed and new theoretical investigations are suggested. 

\end{abstract}




\section{Introduction}

The Crab Nebula has a prominent role in the history of high energy astrophysics.
Ever since its discovery as a supernova 961 years ago, it has exhibited many of the earliest and the best examples of important high energy astrophysical phenomena including synchrotron radiation, spinning, magnetized neutron stars, relativistic outflows and, most recently, dramatic gamma ray flares \citep[e.g.,][]{Hester:2008aa,Buhler:2014aa}.

However, the details of this general interpretation remain controversial. There should be a conversion of electromagnetic to particle energy flux in the outflow and the extent to which this happens, as measured by the magnetization parameter $\sigma\equiv$ Poynting flux/particle energy flux, is quite uncertain \citep[e.g.,][]{Kirk:2009aa,Arons:2012aa}. Another uncertainty is the variation of the wind power with pulsar colatitude, $\theta$. This is generally supposed to decrease from equator to pole but the actual form may depend on the inclination angle of the pulsar magnetic axis \citep{Tchekhovskoy:2015aa}. A third issue is the fate of the magnetic stripes expected to be present in the equatorial outflow \citep{Michel:1971aa}. These have a wavelength of only $10^4$ km, but there may be insufficient time in the outflowing reference frame to dissipate them \citep[e.g.,][]{Coroniti:1990aa, Lyubarsky:2001aa}. A final uncertainty is the nature of the deceleration of the outflow from $\sim c$ to $\sim0.005c$, the nebular expansion speed. It has been commonly presumed that this occurs at a single strong shock with post shock pressure balanced by some ambient nebular pressure \citep{Rees:1974aa,Kennel:1984aa}. However, if $\sigma$ is high and/or the first shock is oblique, the post-shock flow will be high speed or even supersonic and additional shocks may be present. The discovery of the inner knot \citep{Hester:1995aa} was therefore of extreme importance as it promised to address these questions. 

The knot appears as a compact emitting feature located $0.65\arcsec$ southeast of the pulsar. It lies on the rotation axis and is elongated perpendicular to the axis, with a size $0.32\arcsec\times0.16\arcsec$ in optical and $0.40\arcsec\times0.32\arcsec$ in infrared (IR), and is found to be concave towards the pulsar \citep{Rudy:2015aa}. The emission from the knot is highly polarized, with a polarization degree  $\sim60\%$ and position angle approximately along the symmetry axis \citep{Moran:2013aa} . This is consistent with a coherent toroidal magnetic field at the emission site. Meanwhile, the knot emission shows strong variability on a time scale of months to years, though no convincing association with the gamma ray flares \citep{Rudy:2015aa}.

The knot has been convincingly interpreted as Doppler-boosted emission from behind a termination shock \citep{Komissarov:2004aa, Komissarov:2011aa}. However, these new observational results on the compactness, shape and polarization of the knot challenge this interpretation.

In the following we examine the shock model carefully to exhibit these problems, and suggest possible solutions to them. In particular, \S\ref{sec:size} summarizes the problem with the knot size and its offset from the pulsar, \S\ref{sec:curvature} illustrates the curvature problem, \S\ref{sec:depolarization problem} discusses the Doppler-depolarization problem, \S\ref{sec:shock_detail} presents a detailed model of the knot intensity profile and polarization, suggesting additional ingredients needed to fix these problems. In \S\ref{sec:flux} and \S\ref{sec:variability}, we discuss the spectral and temporal features of these observations which provide fresh clues as to the nature of the inner knot. We summarize our conclusions, and suggest some new possible interpretations of the inner knot which point to additional observational and theoretical investigations in \S\ref{sec:discussion}.

After most of the research reported here had been completed, the authors learned of a complementary investigation by Komissarov, Lyutikov and Porth which will be reported elsewhere.

\section{Compactness problem}\label{sec:size}
\begin{figure*}
  \centering
  \includegraphics[width=0.8\textwidth]{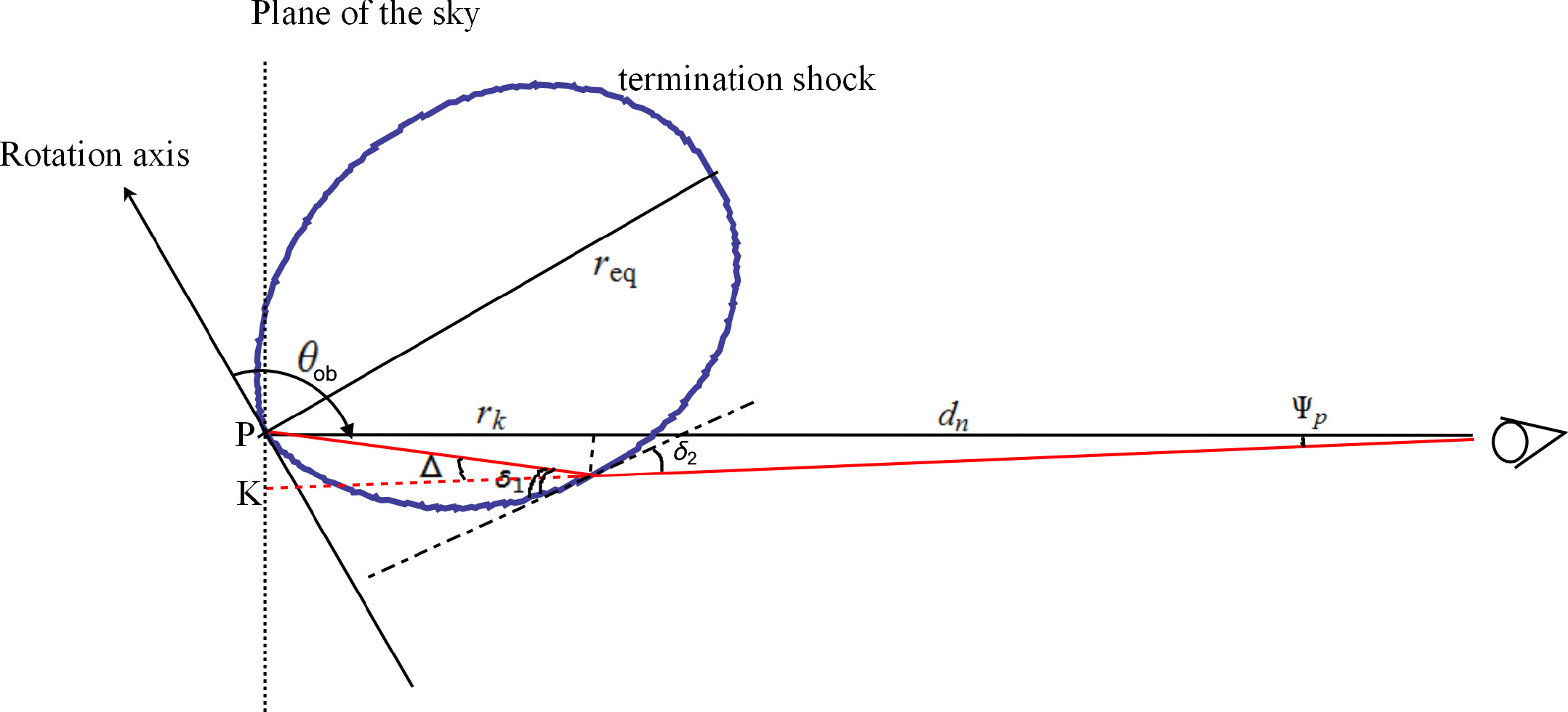}\\
  \caption{Geometry of the termination shock on the meridional plane determined by the pulsar rotation axis and our line of sight toward the pulsar. The pulsar is located at point P. By convention, the pulsar rotation axis, when projected on the plane of the sky (the plane perpendicular to the line of sight, and therefore perpendicular to the plane of the paper), has a position angle $\Psi = 300^{\circ}$ north through east. The angle between line of sight and pulsar rotation axis is $\theta_{\rm ob}=2\pi/3$ \citep[e.g.,][]{Ng:2004aa,Weisskopf:2012aa}. The knot is projected on the plane of the sky at an angle $\Psi_p\approx0.65\arcsec$ southeast of the pulsar (represented by point K here). Other symbols are defined in the text.}\label{fig:shockgeo}
\end{figure*}

Consider a simple adiabatic, ideal MHD shock \citep{Komissarov:2004aa, Komissarov:2011aa} (Figure \ref{fig:shockgeo}). The jump conditions for such a shock assuming an isotropic pressure can be simply obtained by Doppler boosting the jump conditions for a perpendicular shock \citep{Kennel:1984aa} into a frame moving along the shock, or equivalently, following \citet{Komissarov:2011aa} who showed that when the upstream flow is cold and extremely relativistic, the compression ratio across the shock only depends on the upstream magnetization parameter $\sigma\equiv B_1^2/4\pi n_1\gamma_1^2mc^2$, where $B_1$, $n_1$ and $\gamma_1$ are the upstream magnetic field, particle proper density and wind Lorentz factor, respectively (we use subscript 1 for upstream variables and 2 for downstream ones):
\begin{equation}\label{eq:chi}
    \chi\equiv\frac{v_{2n}}{v_{1n}}=\frac{B_1}{B_2}=\frac{1+2\sigma +\sqrt{16\sigma ^2+16\sigma +1}}{6(1+\sigma )}.
\end{equation}
where $v_{1n}$ and $v_{2n}$ are upstream and downstream flow velocity components parallel to the shock normal, respectively. It is easily seen that when $\sigma\to 0$, $\chi\to1/3+4\sigma/3$; when $\sigma\to\infty$, $\chi\to1-1/(2\sigma)$. The downstream bulk Lorentz factor is
\begin{equation}\label{eq:gamma2}
    \gamma _2=\frac{\csc\delta_1}{\sqrt{1-\chi ^2}}.
\end{equation}
where $\delta_1$ is the angle between the upstream velocity and the shock surface. If the shock is highly oblique and/or the magnetization is quite high, the downstream flow would remain relativistic. The emission would then be highly beamed, and the knot corresponds to the spot where the outflow velocity is directed toward us.

\begin{figure}
  \centering
  \includegraphics[width=\columnwidth]{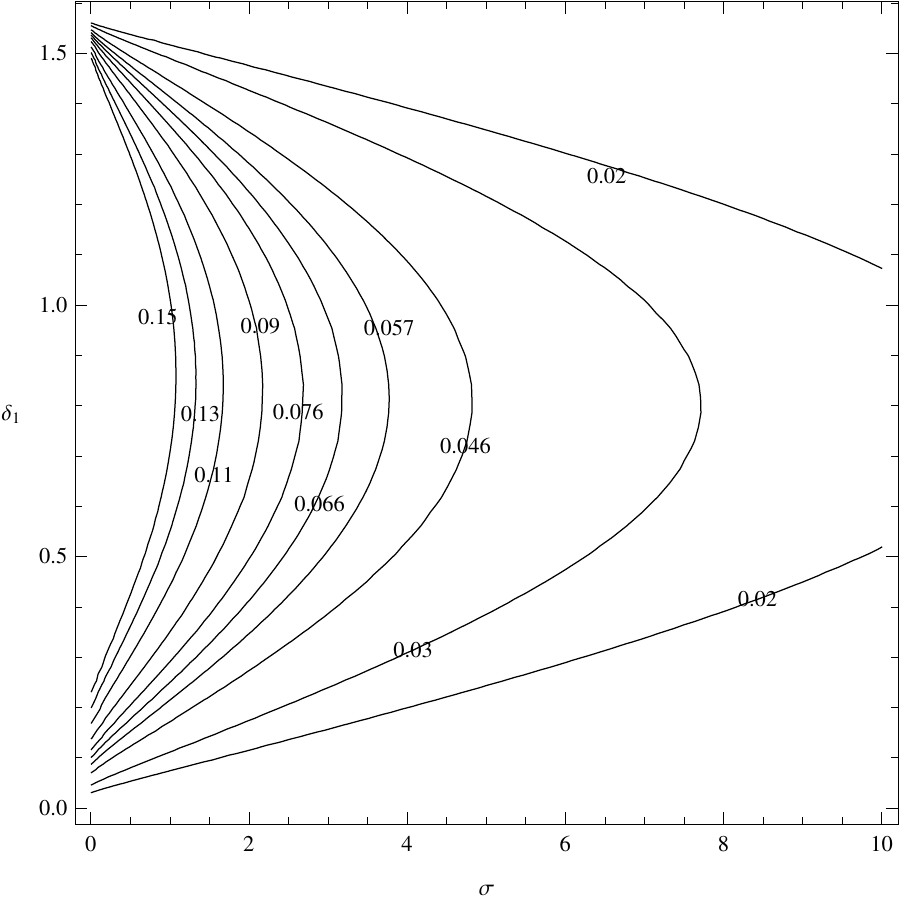}\\
  \caption{Contour plot of deflection angle $\Delta$ on the $\sigma-\delta_1$ plane.}\label{fig:deflection_angle}
\end{figure}

We note that as the shock only reduces the perpendicular velocity, the flow will be deflected from the radial direction. If we denote the angle between the shock surface and downstream velocity as $\delta_2$, it can be shown that
\begin{equation}\label{eq:delta2}
    \tan\delta_2=\chi\tan\delta_1.
\end{equation}
The deflection angle
\begin{equation}\label{eq:deflection_angle}
\Delta=\delta_1-\delta_2=\delta_1-\arctan(\chi\tan\delta_1)
\end{equation}
is thus determined by $\sigma$ and $\delta_1$. The geometry in Figure \ref{fig:shockgeo} indicates that the measured knot-pulsar angular separation should be $\Psi_p=r_k\Delta/D$, where $r_k$ is the actual distance between the pulsar and the emission site at the shock surface, and $D\approx2$ kpc is the distance of the Crab Nebula. Typically $\Psi_p=0.65\arcsec$, and assuming $r_k\le r_{eq}$ where $r_{eq}=11.3\arcsec D$ is the radius of the shock in the equatorial plane \citep{Weisskopf:2012aa}, we get $\Delta\ge0.057$ rad. From the contour plot of $\Delta$ on the $\sigma-\delta_1$ plane (Figure \ref{fig:deflection_angle}), we see that $\sigma$ must be less than 4 for the deflection to be this large.

On the other hand, the knot size perpendicular to the symmetry axis, characterized by its full width at half maximum $\eta_{\perp}$, can be determined from the outflow Lorentz factor, taking into account azimuthal symmetry. Based on a rough estimation that the Doppler beaming angle is $\sim1/\gamma_2$, we have $\eta_{\perp}\approx 2r_k/(\gamma_2 D)$. The ratio $\eta_{\perp}/\Psi_p=2/(\gamma_2\Delta)$ therefore only depends on $\sigma$ and $\delta_1$ and constrains these two parameters. We find that $\eta_{\perp}/\Psi_p=2/(\gamma_2\Delta)=2\sin\delta_1\sqrt{1-\chi^2}/[\delta_1-\arctan(\chi\tan\delta_1)]\ge2.8$, with the minimum happening at $\sigma=0$ and $\delta_1=0.39$. However, $\eta_{\perp}\sim0.5\Psi_p$ is observed \citep{Rudy:2015aa} and so there's clear inconsistency.

The problem remains with more complex shock models, for example, if the upstream plasma is not cold, then as long as $\gamma_1\gg1$, we obtain the same compression ratio as Equation (\ref{eq:chi}) with the definition of $\sigma$ changed into $\sigma=B_1^2/4\pi w_1\gamma_1^2$, where $w_1=\rho_1+P_1$ is the upstream enthalpy density. This does not change $\gamma_2$ or $\Delta$.

Alternatively, if the plasma is anisotropic though still gyrotropic, namely, the particle velocity perpendicular to the local magnetic field is isotropic but differ from parallel velocity distribution, then the compression ratio is given by
\begin{equation}\label{eq:chi_aniso}
\chi=\frac{2+3 \sigma +\zeta  \sigma +\sqrt{(2+5\sigma+3\zeta\sigma)^2+8\sigma(1+\zeta)^2}}{4 (2+\zeta)(1+  \sigma )},
\end{equation}
where $\sigma=B_1^2/4\pi w_1\gamma_1^2=B_1^2/4\pi (\rho_1+P_{1\perp})\gamma_1^2$, $P_{\perp}$ and $P_{\parallel}$ are the pressure tensor components perpendicular and parallel to the local magnetic field, respectively, and we assumed $P_{2\parallel}=\zeta P_{2\perp}$ (for details see Appendix \ref{subsec:aniso_shock}). In the limit $\zeta=0$ (no parallel pressure), we get $\chi =(2+3 \sigma +\sqrt{4+28 \sigma +25 \sigma ^2})/(8+8 \sigma )$. $\chi \to 1/2+3 \sigma /4$ when $\sigma\to0$ and $\chi\to1-2/(5\sigma)$ when $\sigma\to\infty$. The expressions for $\gamma_2$ and $\Delta$ stay the same with $\chi$ changed to this new value, so we get the ratio $\eta_{\perp}/\Psi_p=2/(\gamma_2\Delta)=2\sin\delta_1\sqrt{1-\chi^2}/[\delta_1-\arctan(\chi\tan\delta_1)]\ge3.46$, the minimum occurring when $\sigma=0$ and $\delta_1=0$: this is worse than the isotropic case.  In the limit $\zeta\to\infty$ (no perpendicular pressure) we have $\chi =(\sigma +\sqrt{\sigma  (8+9 \sigma )})/(4 +4\sigma )$. $\chi \to \sqrt{\sigma }/\sqrt{2}+\sigma /4$ as $\sigma\to 0$ and $\chi\to1-2/(3\sigma)$ as $\sigma\to\infty$. In this case, $\eta_{\perp}/\Psi_p=2/(\gamma_2\Delta)\ge1.27$, and the minimum occurs at $\sigma=0$, $\delta_1=\pi/2$. Although parallel pressure dominance could alleviate the tension between the deflection and size, it is difficult to imagine a mechanism that would produce mostly transverse random motion of the particles just behind a shock.

The third possibility is that the upstream flow is a striped wind. In Appendix \ref{subsec:shock_striped} we generalize the shock jump condition of \citet{Lyubarsky:2003aa} to include a possible $\theta$ component of upstream magnetic field and an oblique ahead of shock velocity. We find that the shock jump condition is unchanged if we introduce an effective $\sigma$ that takes into account the reconnection of alternating magnetic flux, in agreement with \citet{Lyubarsky:2003aa}.

\section{Curvature problem}\label{sec:curvature}
\begin{figure*}
  \centering
  \subfigure[]
    {
        \includegraphics[width=0.4\textwidth]{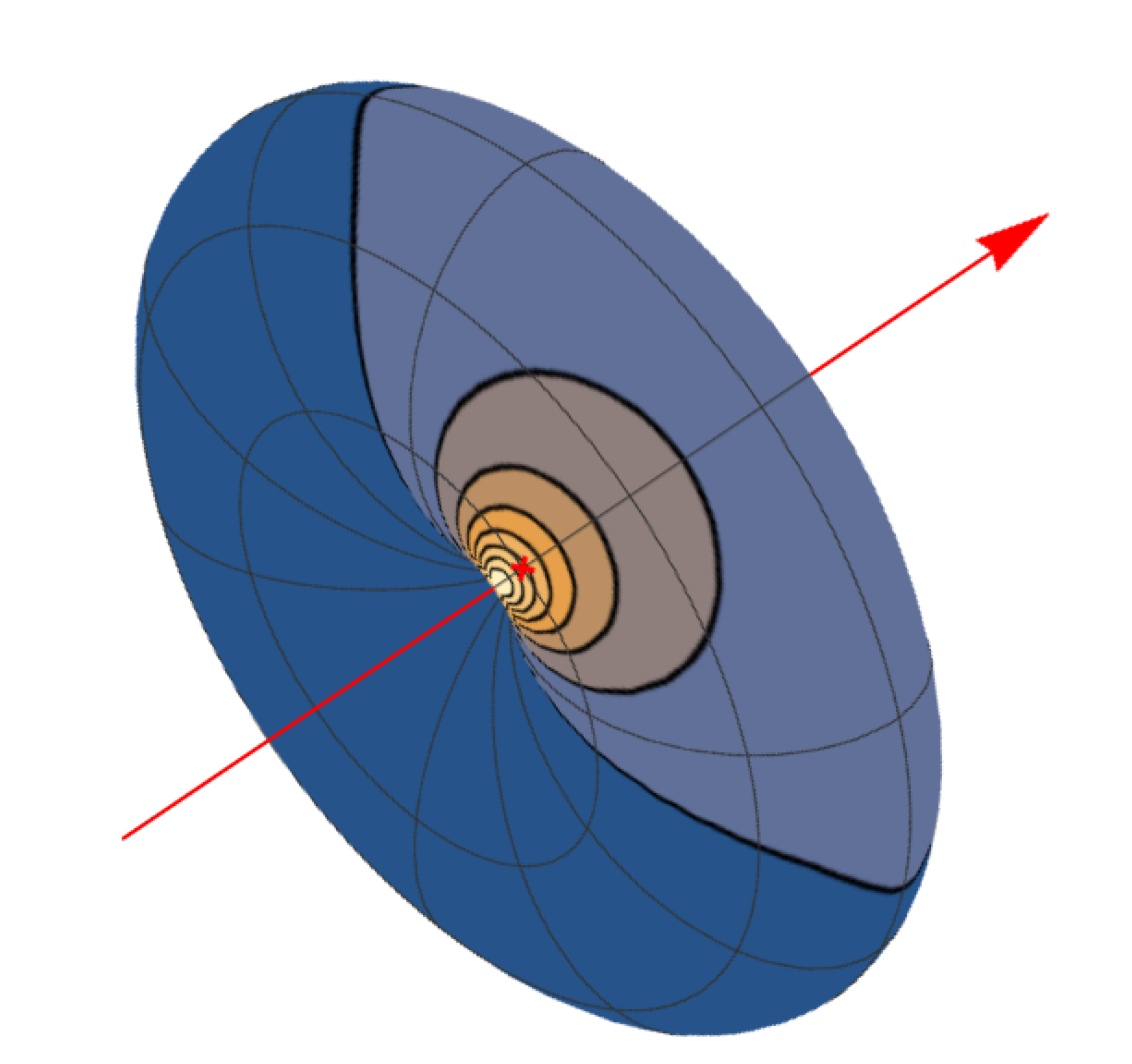}
    }
    \subfigure[]
    {
        \includegraphics[width=0.4\textwidth]{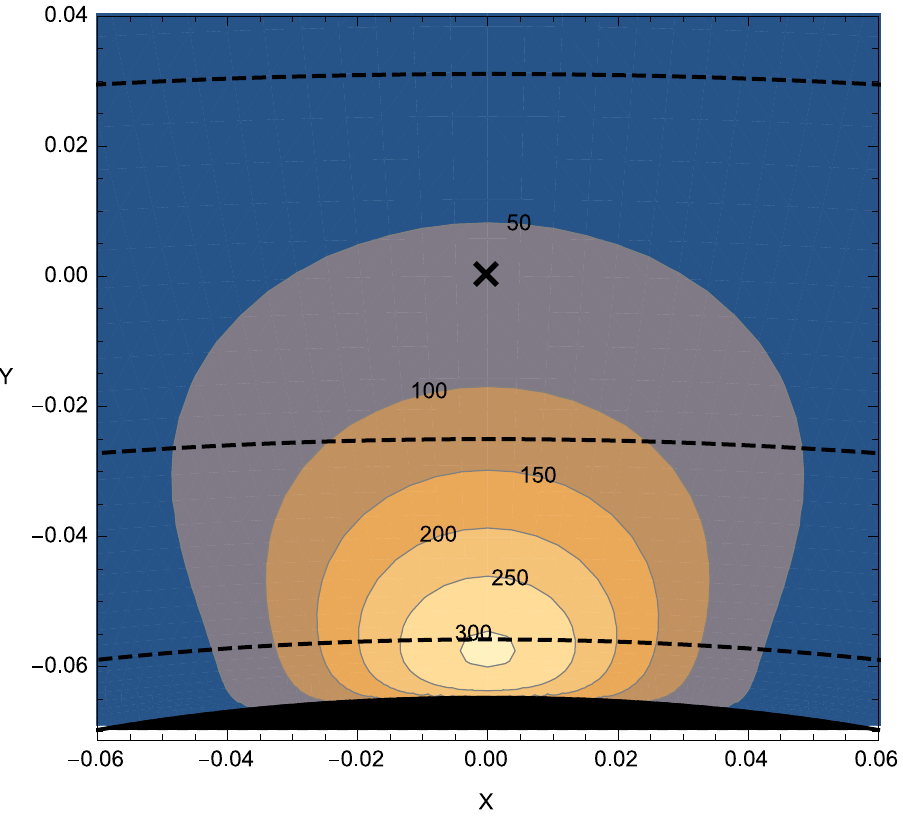}
    }
  \caption{An example showing the emissivity on the surface of the shock. (a) The 3D geometry of the shock surface. In this particular example, we assume the shock surface is given by $r=\sin^4\theta$ and the viewing angle is $\theta_{\rm ob}=2\pi/3$, while the magnetization $\sigma=1$ and the spectral index $\alpha=0.8$. The contours correspond to Doppler factor $\mathscr{D}$ on the shock surface (the innermost contour has $\mathscr{D}=8$ and the interval between successive contours is 1). The location of the pulsar is shown by the red cross and the arrow indicates the pulsar rotation axis. (b) The contour plot of emissivity near the presumed site of the knot, as projected onto the plane of the sky, overlaid with constant $\theta$ curves (black, dashed lines). The location of the pulsar is indicated by the black cross. We assumed that $B\propto1/r$ and particle density $n\propto1/r^2$ ahead of the shock. We used the shock jump condition in Equations (\ref{eq:gamma2})(\ref{eq:deflection_angle}). The procedure of calculating the emissivity is outlined in \S\ref{subsec:emissivity@shock}.}\label{fig:emissivity_shock}
\end{figure*}

In the simple shock model the emission is supposed to arise from a very short length behind the shock and can be approximated as a surface emission. This is illustrated in Figure \ref{fig:emissivity_shock} (a), where the shock surface has a toroidal form. The emissivity at the shock surface is (see \S\ref{sec:shock_detail}) $j_{\nu,\Omega}\propto\mathscr{D}^{2+\alpha}(B'\sin\varpi')^{1+\alpha}$, where $\mathscr{D}=1/[\gamma(1-\vec{\beta}\cdot{\bf n})]=1/[\gamma(1-\beta \cos\psi)]$
is the outflow Doppler factor, $\psi$ is the angle between the line of sight and the flow velocity, $B'$ and $\varpi'$ are the magnetic field and the particle pitch angle in the fluid rest frame, respectively. The $\sin\varpi'$ factor only makes the feature more ``boxy'' (see \S\ref{subsec:intensityMap}) and the curvature is largely determined by the Doppler factor. Assuming axisymmetry, $\gamma$ and $B'$ are only functions of the polar angle $\theta$, and $\psi$ is also dictated by the $\theta$ variation of the flow velocity. (It would be isotropic if we were looking at a spherical surface with radial velocity everywhere.) It is then easily seen that the emissivity will roughly follow the projected constant $\theta$ contours, which are determined by the torus radius and its inclination. (A numerical illustration of the emissivity at a typical shock surface with artificial shape is shown in Figure \ref{fig:emissivity_shock}.) Therefore, if the knot shows any curvature, it should be convex toward the pulsar (a ``frown''). However, \citet{Rudy:2015aa} showed that in both Hubble and Keck images the arc corresponding to the knot is concave towards the pulsar (a ``smile''). Thus, the observed curvature seems to contradict the simple shock model.

\section{Doppler-depolarization Problem}\label{sec:depolarization problem}

If we adopt the best-guess spectral index, $\alpha\sim0.8$ \citep{Sollerman:2003aa,Melatos:2005aa}, the associated degree of linear polarization for a uniform fields is $\sim0.73$ \citep[e.g.,][]{RybickiLightman1986}. This applies in the fluid rest frame. However, we observe the sum of Doppler-boosted photons from different parts of the source. Now when a single photon is aberrated into our line of sight, its electric and magnetic vectors rotate in space about an axis along $\mathbf{k}\times\mathbf{v}$ and the observed electric  vectors will not all be parallel and this contributes a degree of Doppler-depolarization. Typically the integrated degree of polarization will be $\sim0.4-0.5$ \citep[e.g.,][]{Lyutikov:2003aa}. Special conditions are needed to recover the observed polarization of $\sim0.6$. This will be discussed in more detail in \S\ref{subsec:polarization}.

\section{Detailed shock model}\label{sec:shock_detail}

\begin{figure}
  \centering
  \includegraphics[width=\columnwidth]{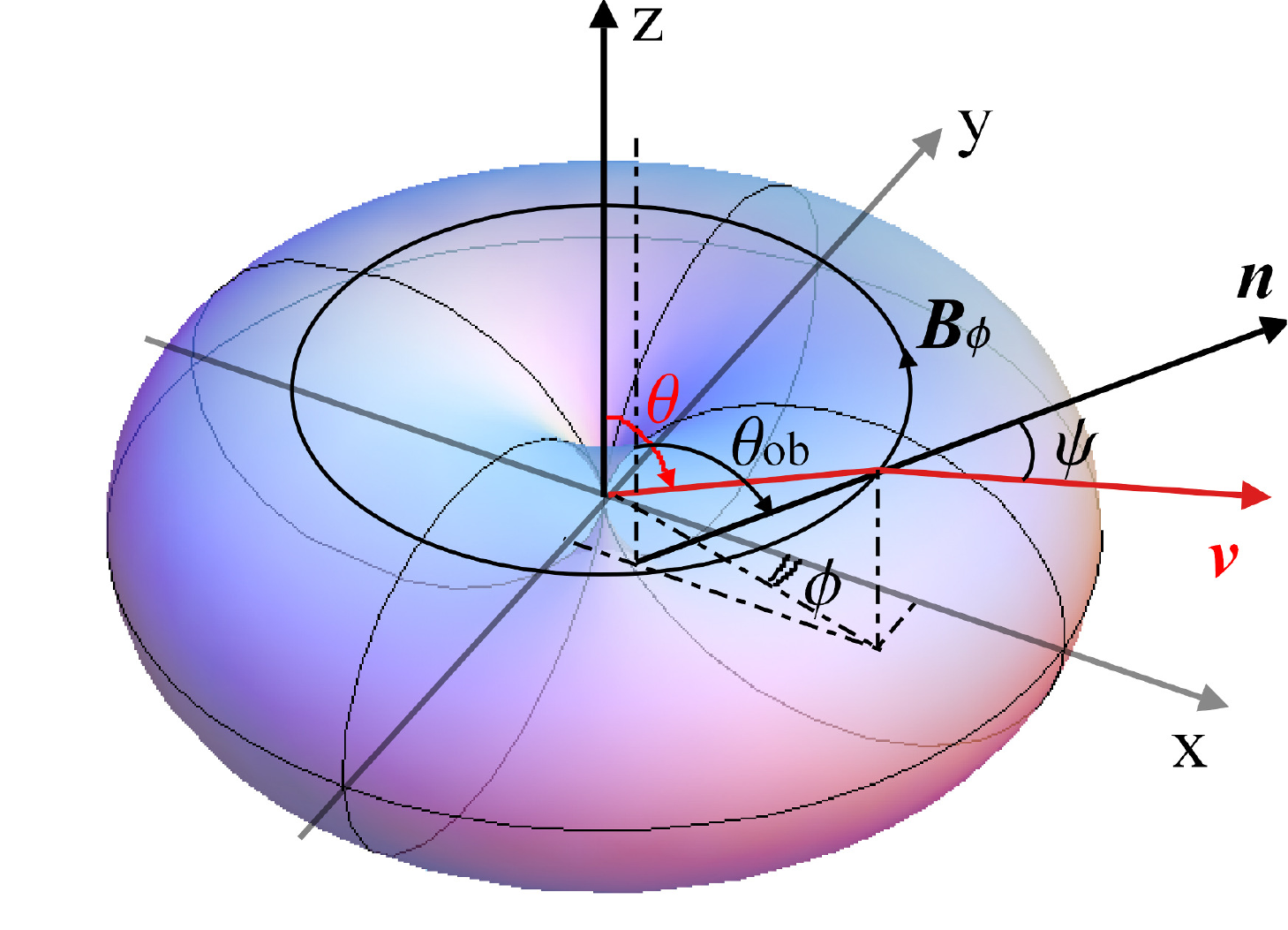}\\
  \caption{Geometry of the shock front. For Crab Nebula the viewing angle should be $\theta_{\rm ob}=2\pi/3$, but here for better presentation we've plotted the case $\theta_{\rm ob}<\pi/2$. The line of sight is along ${\mathbf n}$ and the flow velocity is denoted by ${\mathbf v}$. Magnetic field is toroidal.}\label{fig:shock_geometry}
\end{figure}

Having identified three problems with the simple shock scenario, we now turn to more detailed models to seek ways to address these problems. In this section we scrutinize the emission from the shock and derive the intensity map self-consistently, taking into account post shock flow dynamics. This gives us insights into limitations of the shock model, and suggests what other elements may be needed in a better model.

The shock geometry is shown in Figure \ref{fig:shock_geometry}. Since the knot is quite compact, we anticipate that the post shock flow Lorentz factor is large enough for us to expand about the center of the knot.

We first set up a spherical coordinate system with the $z$ axis along the symmetry axis of the shock (the pulsar rotation axis), and write down the emissivity as a function of the spherical coordinates $\{r,\theta,\phi\}$. The direction of the line of sight, denoted by the unit vector ${\bf n}$, is on the $x-z$ plane, with Cartesian coordinates ${\bf n}=(\sin\theta_{\rm ob},0,\cos\theta_{\rm ob})$. (We use $\{\}$ for spherical coordinates and $()$ for Cartesian coordinates.) For the Crab Nebula, we set $\theta_{\rm ob}=2\pi/3$ \citep[e.g.,][]{Ng:2004aa,Weisskopf:2012aa}. The magnetic field $\vec{B}$ is assumed to lie along the $\hat{\phi}$ direction.

The particle distribution in the fluid rest frame is supposed to be an isotropic, power-law distribution:
\begin{equation}\label{eq:distribution}
f'(\gamma',\varpi')d\gamma' d\Omega' dV'=k'\gamma'^{-(2\alpha+1)}d\gamma' d\Omega' dV',
\end{equation}
where $\varpi'$ is the pitch angle of the particles in the fluid rest frame. If the outflow bulk Lorentz factor is $\gamma$, the emissivity of the outflow would be
\begin{equation}\label{eq:emissivity}
\begin{split}
    j_{\nu,\Omega}(\nu,\mathbf{n})&=\mathscr{D}^2j'_{\nu',\Omega'}(\nu'=\nu/\mathscr{D},\mathbf{n}')= \mathscr{D}^{2+\alpha}j'_{\nu',\Omega'}(\nu,\mathbf{n}')\\
    &=C(\nu) k'\mathscr{D}^{2+\alpha}(B'\sin\varpi')^{1+\alpha},
\end{split}
\end{equation}
where
\begin{equation}
\mathscr{D}=1/[\gamma(1-\vec{\beta}\cdot{\bf n})]=1/[\gamma(1-\beta \cos\psi)]
\end{equation}
is the outflow Doppler factor, $\psi$ is the angle between the line of sight and the flow velocity, and
\begin{equation}
C(\nu)=\frac{\alpha+5/3}{\alpha+1}\,\frac{\sqrt{3}}{4}\, \Gamma\!\left(\frac{\alpha}{2}+\frac{1}{6}\right) \Gamma\!\left(\frac{\alpha}{2}+\frac{5}{6}\right)\frac{e^3}{mc^2} \left(\frac{2\pi mc}{3e}\right)^{-\alpha}\nu^{-\alpha}.
\end{equation}
The intensity is then
\begin{equation}\label{eq:intensity}
I_{\nu,\Omega}(\nu,\mathbf{n})=\int j_{\nu,\Omega}(\nu,\mathbf{n})ds,
\end{equation}
where the integral is carried out along the line of sight. (We omit the subscripts $\nu$ and $\Omega$ on $j$ or $I$ in the following calculations but readers should keep in mind that these refer to quantities per steradian and per unit frequency.)

\subsection{Post shock flow dynamics and emissivity}\label{subsec:flowdynamics}
The flow should be axisymmetric and poloidal. If we consider a small bundle of flow lines that has a cross section $\delta A=\delta\ell\cdot r\delta\phi$, where $\delta\ell$ is the width measured in the poloidal plane and $r\delta\phi$ is that measured in the toroidal direction, for a steady flow we have the following conservation laws:
\begin{itemize}
  \item Conservation of particle number: $nu\cdot\delta A=const$;
  \item Conservation of energy: $[(nm+\frac{\Gamma}{\Gamma-1}P)\gamma u+\frac{v B^2}{4\pi }]\delta A=const$;
  \item Conservation of magnetic flux (or $\nabla\times{\mathbf E}=0$): $vb\cdot\delta\ell=const$;
  \item Equation of state: $P=kn^{\Gamma}$, where $k$ is a constant for this flow line.
\end{itemize}
If the outflow were spherically symmetric and the post shock Lorentz factor is sufficiently high, $B'\propto1/ur$ and $n\propto1/ur^2$, and if the particles evolve adiabatically, $k'\propto n^{2\alpha/3+1}$, so
\begin{equation}
j\propto\left[\frac{1}{\gamma(1-\beta\cos\psi)}\right]^{2+\alpha}u^{-(5\alpha/3+2)}r^{-(7\alpha/3+3)}.
\end{equation}
In the large $\gamma$ limit, we have
\begin{equation}
j\propto(1+\gamma^2\psi^2)^{-(2+\alpha)}\beta^{-(2+\alpha)}u^{-2\alpha/3}r^{-(7\alpha/3+3)}
\end{equation}
and the line integral in Equation (\ref{eq:intensity}) would be dominated by emissivity close to the shock surface. We consider the influence of curvature in the flow below.

\subsection{Emissivity at the shock front}\label{subsec:emissivity@shock}
Suppose that the shock surface is a smooth function of $\theta$: $r=r(\theta)$. The radial wind that goes through the point $\{r(\theta),\theta,\phi\}$ on the shock surface is deflected by an angle $\Delta$ and now is directed along a different polar angle $\theta_2$. Then the angle $\psi$ between ${\bf n}$ and $\vec{\beta}_2$ is \citep[e.g., ][]{Lyutikov:2003aa}
\begin{equation}\label{eq:cospsi}
    \cos\psi=\cos\theta_{\rm ob}\cos{\theta_2}+\sin\theta_{\rm ob}\sin\theta_2\cos\phi.
\end{equation}
Since the radiation is mostly concentrated along the line of sight, i.e. $|\theta_2-\theta_{\rm ob}|\ll1$, we can expand around the flow line that is directed toward us. To lowest order approximation, we have
\begin{equation}\label{eq:cospsi1}
    \cos\psi\approx1-\frac{1}{2}d\theta_2^2-\frac{1}{2}\sin^2\theta_{\rm ob}\phi^2.
\end{equation}
A photon propagating along the direction ${\bf n}$ in the nebula frame is emitted along the direction of ${\bf n'}$ in the fluid rest frame \citep[e.g., ][]{Lyutikov:2003aa}:
\begin{equation}
{\bf n'}=\frac{{\bf n}+\gamma_2\vec{\beta}_2(\frac{\gamma_2}{\gamma_2+1}{\bf n}\cdot\vec{\beta}-1)}{\gamma_2(1-{\bf n}\cdot\vec{\beta}_2)}.
\end{equation}
Since the magnetic field in the fluid rest frame is along $\hat{\phi}$ direction, $\hat{B'}=(-\sin\phi,\cos\phi,0)$, the angle between the photon and $\hat{B'}$ is \citep[e.g., ][]{Lyutikov:2003aa}:
\begin{equation}
\cos\varpi'={\bf n'}\cdot\hat{B'}=\frac{{\bf n}\cdot\hat{B'}}{\gamma_2(1-{\bf n}\cdot\vec{\beta}_2)}=\frac{-\sin\theta_{\rm ob}\sin\phi}{\gamma_2(1-{\bf n}\cdot\vec{\beta}_2)},
\end{equation}
so
\begin{equation}
\sin\varpi'=\left[1-\frac{\sin^2\theta_{\rm ob}\sin^2\phi}{\gamma_2^2(1-\beta_2\cos\psi)^2}\right]^{1/2}.
\end{equation}

In the following we derive $\theta_2$ as a function of $\theta$. At a point $\{r(\theta),\theta,\phi\}$ on the shock surface, the incident angle $\delta_1$ is
\begin{equation}\label{eq:incident_angle}
    \cot\delta_1=\frac{1}{r}\frac{dr}{d\theta},
\end{equation}
and the curvature of the shock in the meridional plane is
\begin{equation}\label{eq:curvature}
    \kappa=\frac{r^2+2\left(\frac{dr}{d\theta}\right)^2-r\frac{d^2r}{d\theta^2}}{\left[\left(\frac{dr}{d\theta}\right)^2+r^2\right]^{3/2}}.
\end{equation}
According to the shock jump condition, the deflection angle is
\begin{equation}
    \Delta=\delta_1-\arctan(\chi\tan\delta_1),
\end{equation}
where $\chi$ is the compression ratio. Consider the flow line that is directed toward us: $\theta_{2}=\theta_{\rm ob}$; writing this point as $\{r(\theta_0),\theta_0,0\}$, we have $\theta_0+\Delta_0=\theta_{\rm ob}$. Now if we deviate from this point by a small polar angle $d\theta$, we find that $d\theta_2=d\theta+d\Delta$, and since
\begin{equation}
    \frac{d\Delta}{d\delta_1}=1-\frac{\chi\sec^2\delta_1}{1+\chi^2\tan^2\delta_1},
\end{equation}
\begin{equation}
    \frac{d\delta_1}{d\theta}=\frac{\left(\frac{dr}{d\theta}\right)^2-r\frac{d^2r}{d\theta^2}}{r^2+\left(\frac{dr}{d\theta}\right)^2},
\end{equation}
\begin{equation}
    \frac{d\Delta}{d\theta}=\frac{d\Delta}{d\delta_1}\frac{d\delta_1}{d\theta},
\end{equation}
we get
\begin{equation}\label{eq:dtheta}
    d\theta_2=d\theta\left(\chi\frac{\cos^2\delta_2}{\cos^2\delta_1}+(1-\chi\frac{\cos^2\delta_2}{\cos^2\delta_1})\frac{\kappa r_k}{\sin\delta_1}\right)\equiv a\,d\theta,
\end{equation}
where 
\begin{equation}
a=\chi\frac{\cos^2\delta_2}{\cos^2\delta_1}+\left(1-\chi\frac{\cos^2\delta_2}{\cos^2\delta_1}\right)\frac{r_k}{R_c\sin\delta_1}.
\end{equation}

\begin{figure}
  \centering
  \includegraphics[width=\columnwidth]{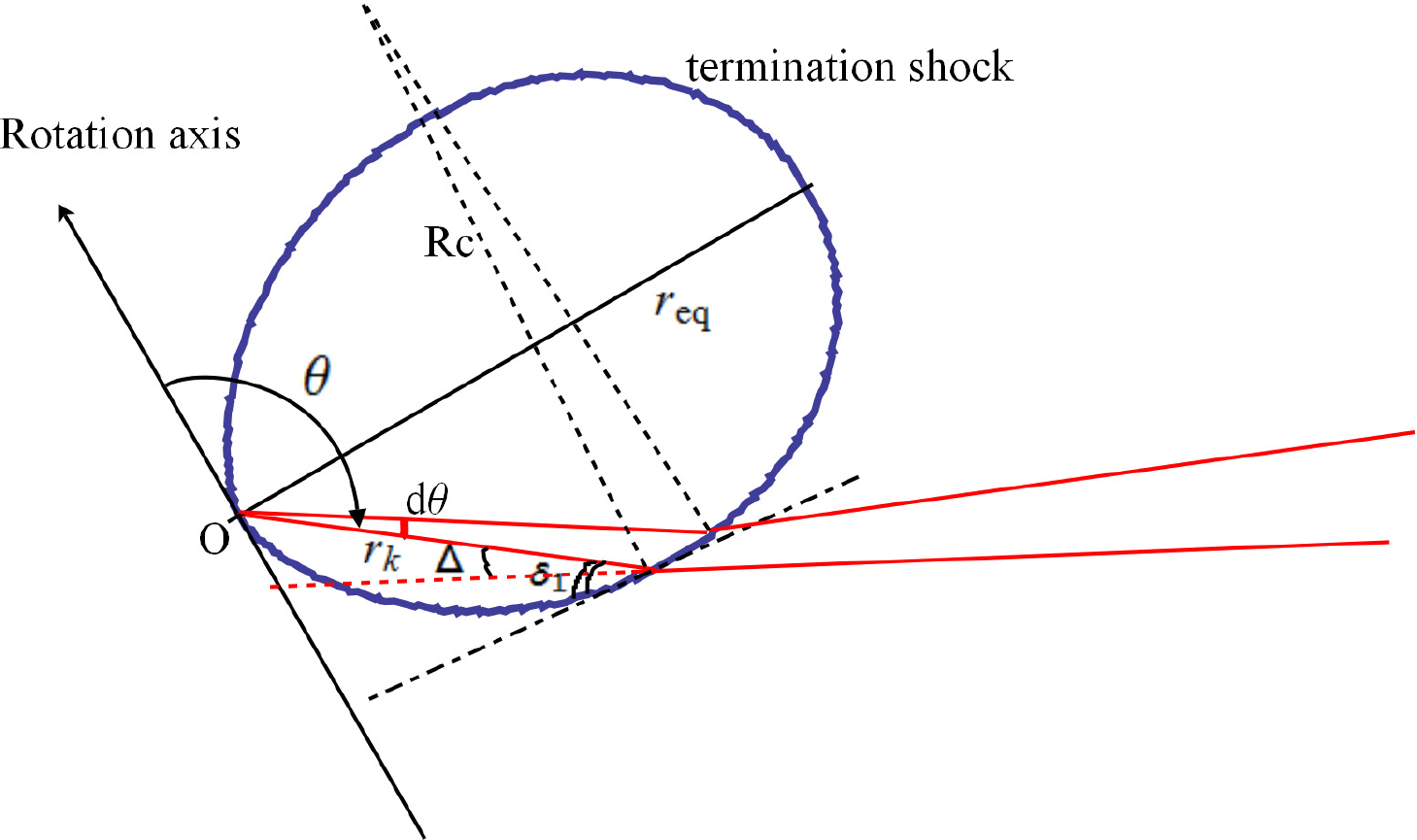}\\
  \caption{A meridional cross section of the shock.}\label{fig:shock_meridional}
\end{figure}

The same result can be derived based on a geometrical argument, see Figure \ref{fig:shock_meridional}. Suppose that the shock radius of curvature in the meridional plane is $R_c$. Consider a small portion of the incoming wind with an opening angle $d\theta$, the difference in upstream incident angle is 
\begin{equation}
d\delta_1=-d\theta+d\theta \kappa r_k/\sin\delta_1.
\end{equation}
Since the shock jump condition gives $\tan\delta_2=\chi\tan\delta_1,$
we have
\begin{equation}
d\delta_2=d\delta_1\chi\cos^2\delta_2/\cos^2\delta_1.
\end{equation}
Thus the outflow opening angle is
\begin{equation}
\begin{split}
d\theta_2&=-d\delta_2+d\theta r_k/(R_c \sin\delta_1) \\
&=d\theta\left(\chi\frac{\cos^2\delta_2}{\cos^2\delta_1}+(1-\chi\frac{\cos^2\delta_2}{\cos^2\delta_1})\frac{r_k}{R_c\sin\delta_1}\right)=a\,d\theta.
\end{split}
\end{equation}
\begin{figure}
  \centering
  \includegraphics[width=0.5\columnwidth]{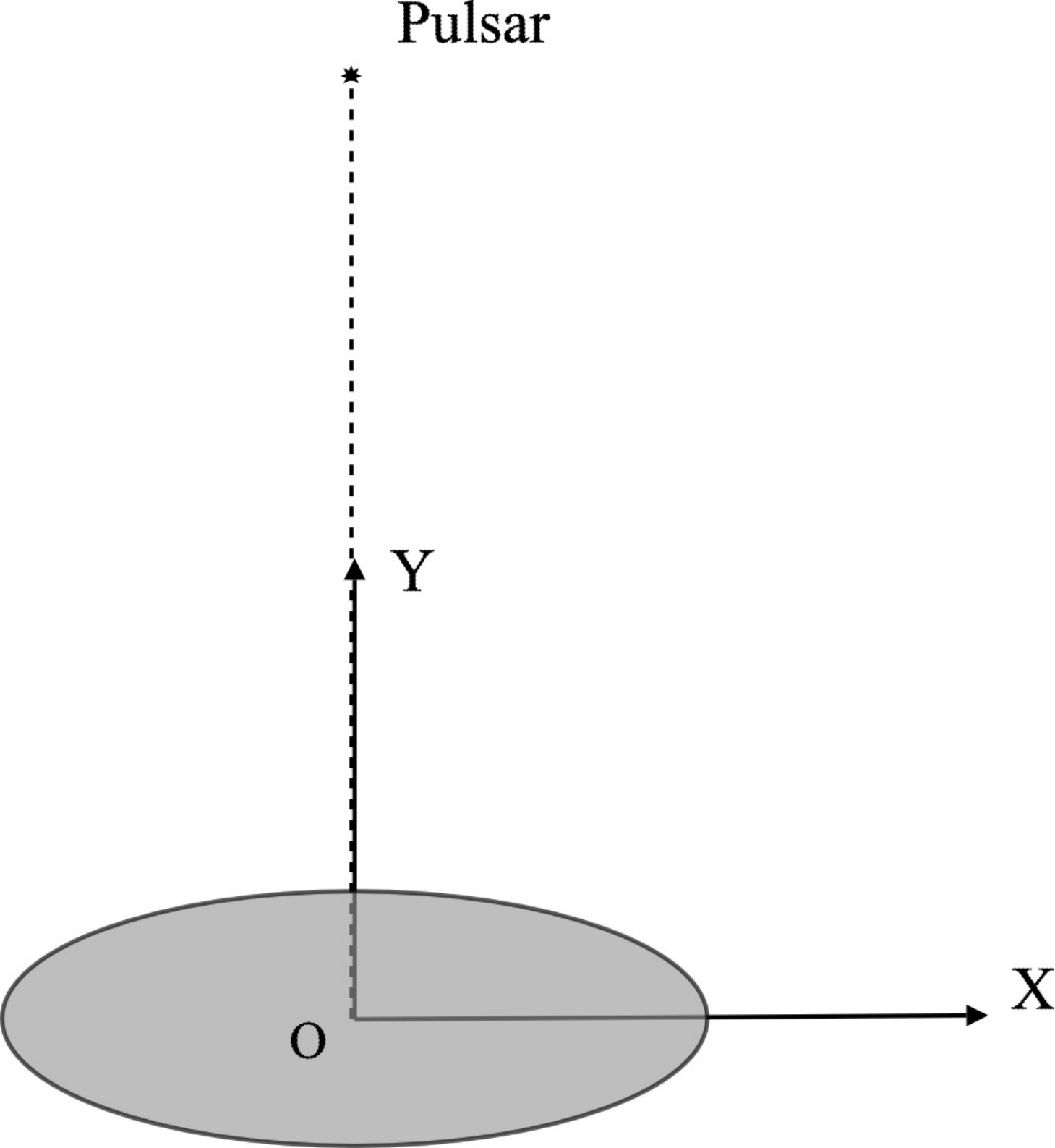}\\
  \caption{Coordinate system for the projected intensity map on the plane of the sky. The origin is located at the peak intensity point of the knot, corresponding to the flow line behind the shock that is directed toward us. The pulsar is located somewhere on the positive Y axis.}\label{fig:projection_coordinates}
\end{figure}

Now consider the projection. If we expand around the flow line that is directed toward us, namely the point $\{r(\theta_0),\theta_0,0\}$ in spherical coordinates or $(x_0,0,z_0)$ in Cartesian coordinates, we have
\begin{equation}
\begin{aligned}
  dx &= \left.\left(\frac{dr}{d\theta}\sin\theta+r\cos\theta\right)\right|_{\theta=\theta_0}d\theta, \\
  dy &= r_k\sin\theta_0d\phi, \\
  dz &= \left.\left(\frac{dr}{d\theta}\cos\theta-r\sin\theta\right)\right|_{\theta=\theta_0}d\theta.
\end{aligned}
\end{equation}
The coordinate system we use for the image on the plane of the sky is shown in Figure \ref{fig:projection_coordinates}. The projected coordinates $(X, Y)$ corresponds to a rotation of the original pulsar-based $x,\,y,\,z$ coordinates
\begin{equation}
\begin{aligned}
  dX &= dy, \\
  dY &= -\cos\theta_{\rm ob}dx+\sin\theta_{\rm ob}dz.
\end{aligned}
\end{equation}
Since $\theta_0+\Delta_0=\theta_{\rm ob}$, we finally get
\begin{equation}
\begin{aligned}
  X &= r_k\sin(\theta_{\rm ob}-\Delta_0)d\phi, \\
  Y &= -r_k(\cos\Delta_0-\cot\delta_1\sin\Delta_0)d\theta,
\end{aligned}
\end{equation}
near the point where the flow line is directed toward us.

Using all the relations derived above, assuming $\gamma_2\gg1$, neglecting variation of $\gamma_2$, $B'$ and $k'$ over the emitting region, we have, to lowest order of $X$, $Y$,
\begin{equation}
\mathscr{D}\approx\frac{2\gamma_2}{1+\gamma_2^2\psi^2}\approx\frac{2\gamma_2}{1+\gamma_2^2X^2/x_0^2+a^2\gamma_2^2Y^2/y_0^2},
\end{equation}
\begin{equation}
\begin{split}
\sin\varpi'&\approx\left[1-\frac{4\sin^2\theta_{\rm ob}\gamma_2^2\phi^2}{(1+\gamma_2^2\psi^2)^2}\right]^{1/2}\\
&\approx\left[1-\frac{4\gamma_2^2X^2/x_0^2}{(1+\gamma_2^2X^2/x_0^2+a^2\gamma_2^2Y^2/y_0^2)^2}\right]^{1/2},
\end{split}
\end{equation}
\begin{multline}\label{eq:emissivity_shock}
j(X,Y)\approx C(\nu) k' B'^{1+\alpha} \left[\frac{2\gamma_2}{1+\gamma_2^2\psi^2}\right]^{2+\alpha} \left[1-\frac{4\sin^2\theta_{\rm ob}\gamma_2^2\phi^2}{(1+\gamma_2^2\psi^2)^2}\right]^{(1+\alpha)/2}\\
\approx C(\nu) k' B'^{1+\alpha} \left[\frac{2\gamma_2}{1+\gamma_2^2X^2/x_0^2+a^2\gamma_2^2Y^2/y_0^2}\right]^{2+\alpha}\\
 \times\left[1-\frac{4\gamma_2^2X^2/x_0^2}{(1+\gamma_2^2X^2/x_0^2+a^2\gamma_2^2Y^2/y_0^2)^2}\right]^{\frac{1}{2}(1+\alpha)},
\end{multline}
where $x_0=r_k\sin(\theta_{\rm ob}-\Delta_0)/\sin\theta_{\rm ob}$, $y_0=r_k(\cos\Delta_0-\cot\delta_1\sin\Delta_0)$. If $\Delta_0\ll1$, $x_0\approx r_k$, $y_0\approx r_k(1-\cot\delta_1\tan\Delta)$; further more if the upstream velocity is quasi-parallel to the shock surface, namely $\delta_1\rightarrow0$, we would have $y_0\rightarrow \chi r_k\approx\chi x_0$, but if the flow is quasi-perpendicular ($\delta_1\rightarrow\pi/2$), we have $y_0\approx r_k\approx x_0$. This already hints that we are more likely to get a knot that is elongated perpendicular to the symmetry axis when the upstream flow is quasi-parallel and $\sigma$ is small as we now discuss.

\subsection{Intensity map}\label{subsec:intensityMap}
More generally, the emissivity in the outflow is (in the following we omit the subscript ``2'' and just use $\gamma$, $\beta$, $u$ to denote the outflow Lorentz factor, velocity and proper velocity, respectively, not necessarily at the shock front)
\begin{equation}
j\approx C(\nu) k' B'^{1+\alpha} \left[\frac{2\gamma}{1+\gamma^2\psi^2}\right]^{2+\alpha} \left[1-\frac{4\sin^2\theta_{\rm ob}\gamma^2\phi^2}{(1+\gamma^2\psi^2)^2}\right]^{(1+\alpha)/2}.
\end{equation}
Now supposing that the outflow roughly follows a spherical expansion, we have $B'\propto1/ur$ and $n\propto1/ur^2$, and if the particles evolve adiabatically, $k'\propto n^{2\alpha/3+1}$, so
\begin{multline}\label{eq:emissivity_general}
j\approx C_1(\nu) u^{-2\alpha/3}\beta^{-(2+\alpha)}r^{-7\alpha/3-3} (1+\gamma^2\psi^2)^{-(2+\alpha)} \\
\times\left[1-\frac{4\sin^2\theta_{\rm ob}\gamma^2\phi^2}{(1+\gamma^2\psi^2)^2}\right]^{(1+\alpha)/2},
\end{multline}
where the constants have been absorbed into the new prefactor $C_1(\nu)$.

The outflow may bend in poloidal direction. We introduce a parameter $\tau=r_k\,d\theta/dr$ to characterize this bending, so that at a radius $r$, the direction of the outflow has a polar angle
\begin{equation}\label{eq:bending}
\theta=\theta_{2s}+\tau(r/r_k-1),
\end{equation}
where $\theta_{2s}$ is the velocity polar angle at the shock front. For simplicity, suppose $\tau$ is more or less the same over the emitting region.

If the outflow is not curved, then from the geometry, we can see that the angle $\psi$ between the line of sight and the velocity at radius $r$ satisfies (at small angle limit, valid when $a>0$, namely the deflection is monotonic in the emitting patch)
\begin{equation}
\begin{split}
\psi^2&\approx\frac{r_k^2}{r^2}\sin^2\theta_{\rm ob}\phi_s^2+\left(\frac{r_k/a}{r-r_k+r_k/a}\right)^2(\theta_{2s}-\theta_{\rm ob})^2\\
&\approx\frac{r_k^2}{r^2}\sin^2\theta_{\rm ob}\phi_s^2+\left(\frac{r_k}{r-r_k+r_k/a}\right)^2(\theta_{s}-\theta_0)^2\\
&\approx\frac{X^2r_k^2}{x_0^2r^2}+\frac{Y^2r_k^2}{y_0^2(r-r_k+r_k/a)^2},
\end{split}
\end{equation}
where subscript s denotes quantities at the shock front. Based on this, we can easily see that if the flow bends according to Equation (\ref{eq:bending}), then
\begin{multline}\label{eq:psi2}
    \psi^2\approx\frac{r_k^2}{r^2}\sin^2\theta_{\rm ob}\phi^2+\left(\frac{r_k}{r-r_k+r_k/a}(\theta_{s}-\theta_0)+\tau(r/r_k-1)\right)^2\\
    \approx\frac{X^2r_k^2}{x_0^2r^2}+\left(-\frac{Yr_k}{y_0 (r-r_k+r_k/a)}+\tau(r/r_k-1)\right)^2.
\end{multline}
$\tau>0$ means bending toward the pole; $\tau<0$ means bending toward the equator; $\tau=0$ takes us back to the straight flow.

Parameterizing the distance along the line of sight as $s$, we get the intensity from the line integral
\begin{multline}\label{eq:I_integral}
    I(\nu,\mathbf{n})\approx \int ds j(\nu,\mathbf{n})
    \approx\int\frac{dr}{\hat{r}\cdot\hat{n}}j(\nu,\mathbf{n})\\
    \approx C_1(\nu)\int_{r_k}^{\infty} dr\ u^{-2\alpha/3}\beta^{-(2+\alpha)}r^{-7\alpha/3-3} (1+\gamma^2\psi^2)^{-(2+\alpha)} \\
    \times\left[1-\frac{4\sin^2\theta_{\rm ob}\gamma^2\phi^2}{(1+\gamma^2\psi^2)^2}\right]^{(1+\alpha)/2},
\end{multline}
where $\psi^2$ is determined by Equation (\ref{eq:psi2}). In the last line $\hat{r}\cdot\hat{n}$ is taken to be 1 because $\psi\sim O(1/\gamma)$. The upper limit on the integral can be set to $\infty$ since the emissivity is only significant close to the shock.

Let's look at a few examples. In the following we assume that the flow does not slow down rapidly so the change of $u$, $\gamma$ and $\beta$ over a short distance can be neglected. The integral in Equation (\ref{eq:I_integral}) can then be carried out numerically.

\subsubsection{Straight outflow}

\begin{figure*}
  \centering
  \subfigure[]
    {
        \includegraphics[width=0.4\textwidth]{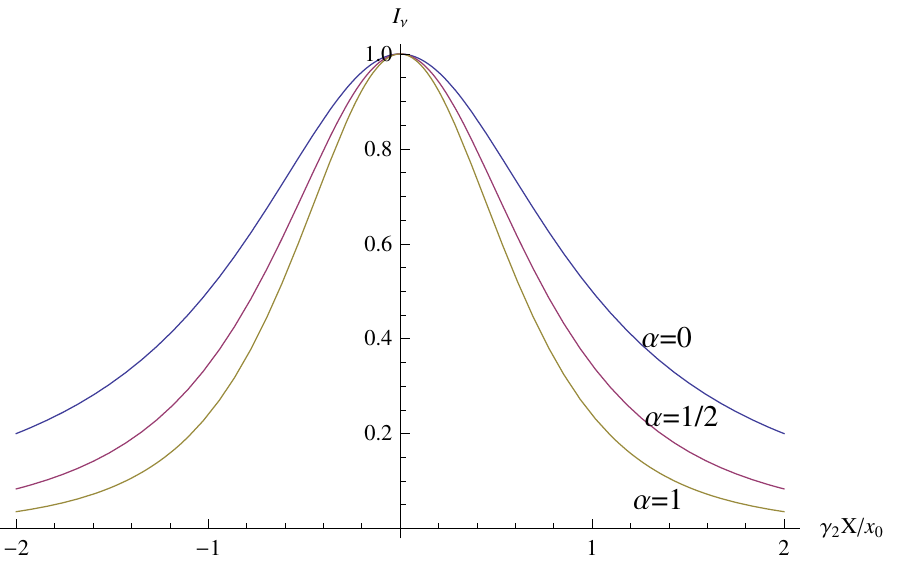}
    }
    \subfigure[]
    {
        \includegraphics[width=0.4\textwidth]{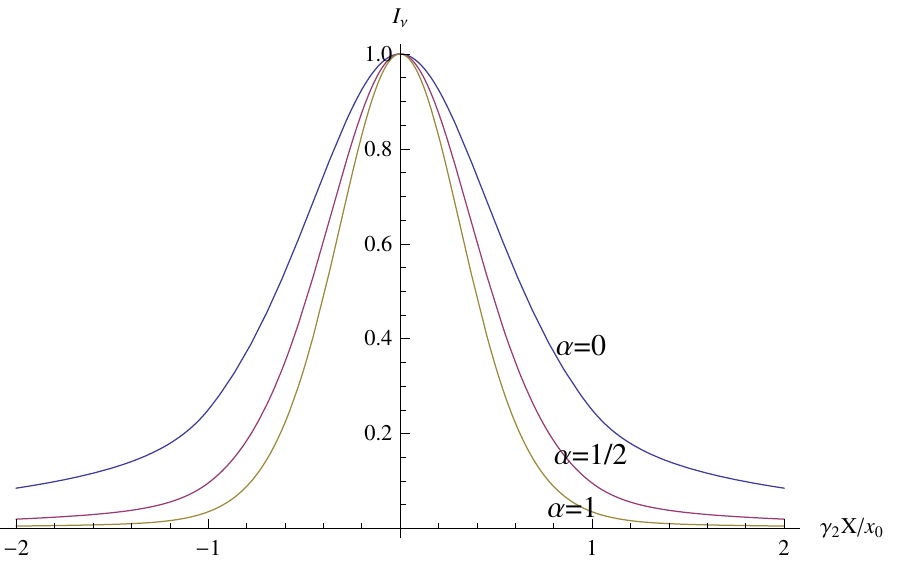}
    }
  \caption{(a) Intensity profile along the line $Y=0$ for different spectral indices $\alpha$=0, 0.5, 1, when the outflow is straight and the pitch angle dependence is not taken into account. The profiles are normalized at the peak intensity. (b) Same as (a), except that the pitch angle dependence of synchrotron power is included.}\label{fig:straightIprofile}
\end{figure*}
\begin{figure*}
    \centering
    \subfigure[]
    {
        \includegraphics[width=0.3\textwidth]{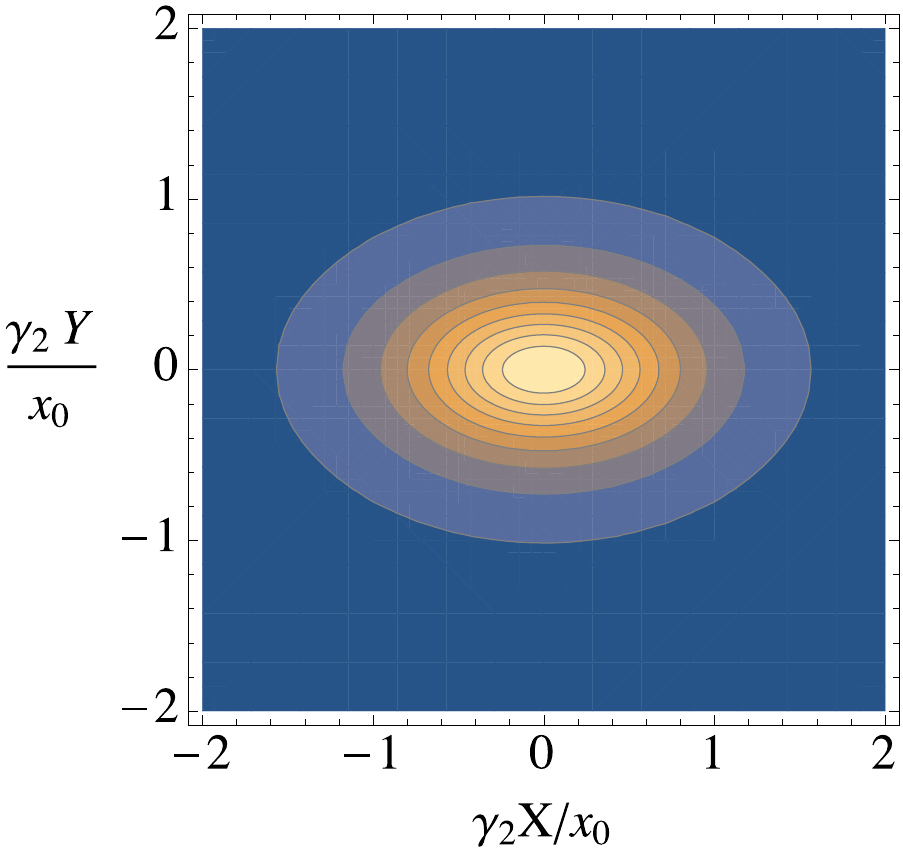}
    }
    \subfigure[]
    {
        \includegraphics[width=0.3\textwidth]{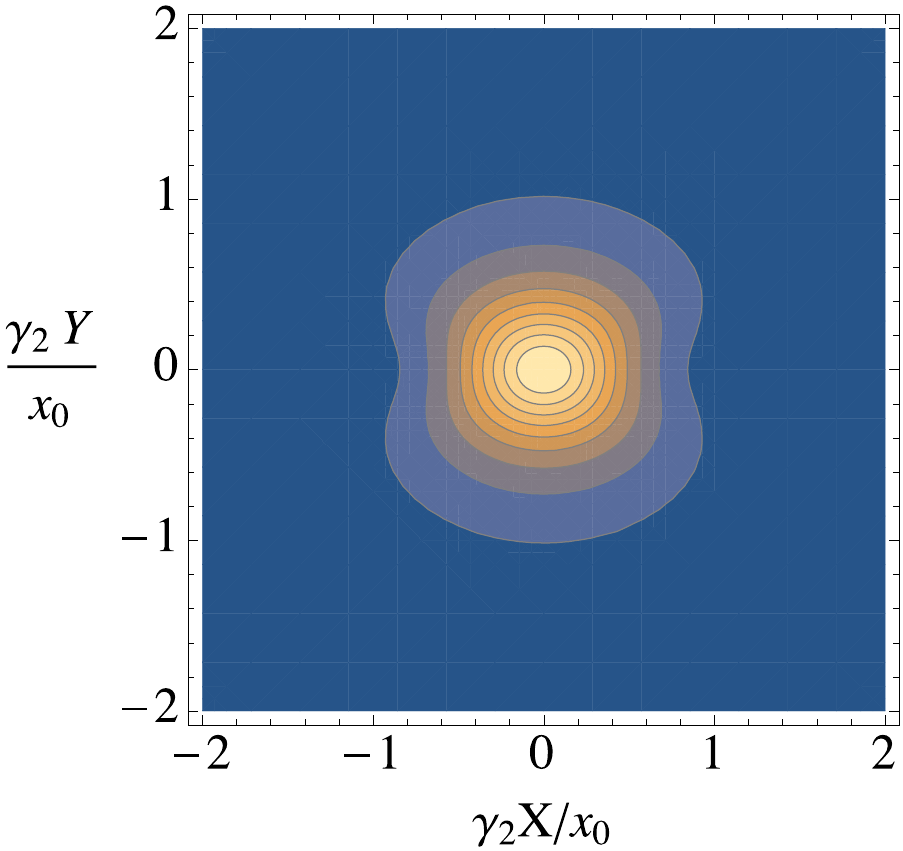}
    }
     \subfigure[]
    {
        \includegraphics[width=0.3\textwidth]{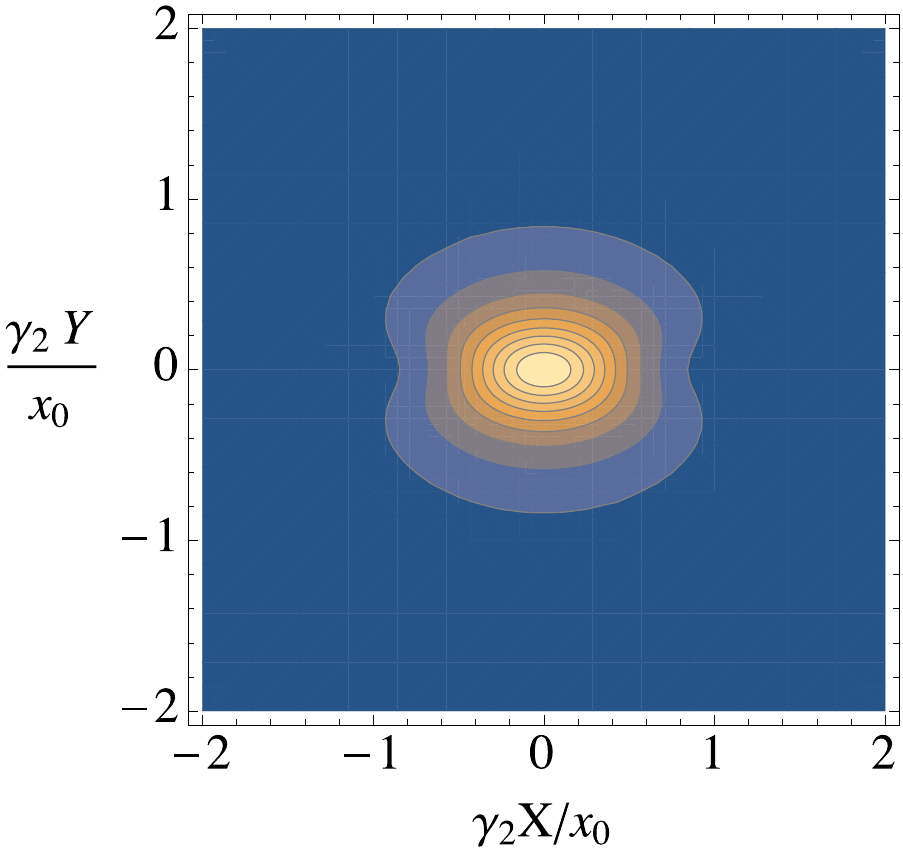}
    }
    \caption{Knot shape and intensity profile for the case of straight outflow, in the coordinate system of Figure \ref{fig:projection_coordinates}. (a) Contour Plot for a special example with aspect ratio parameter $b=2$ and spectral index $\alpha=0.8$, when the pitch angle dependence is neglected; (b) Same as (a), except that the pitch angle dependence of synchrotron power is included: the knot gets squeezed in the x direction; (c) same as (b), except that $b=3$. The intensity is normalized by the peak value and the contours go from 0.1 to 0.9 with increment 0.1. Same below.}\label{fig:straightI}
\end{figure*}

\begin{figure}
    \centering
    \subfigure[]
    {
        \includegraphics[width=\columnwidth]{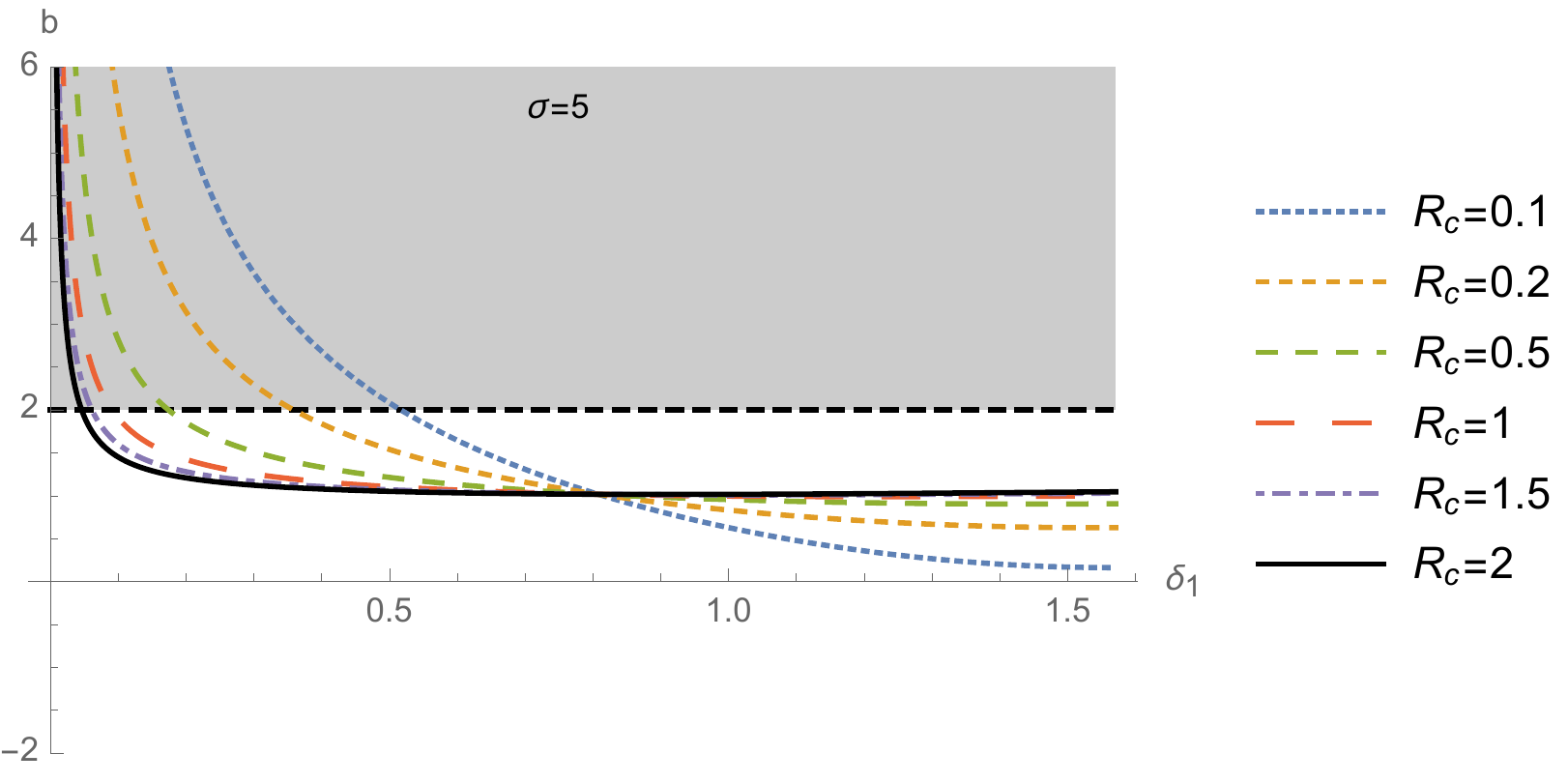}
    }\\
    \subfigure[]
    {
        \includegraphics[width=\columnwidth]{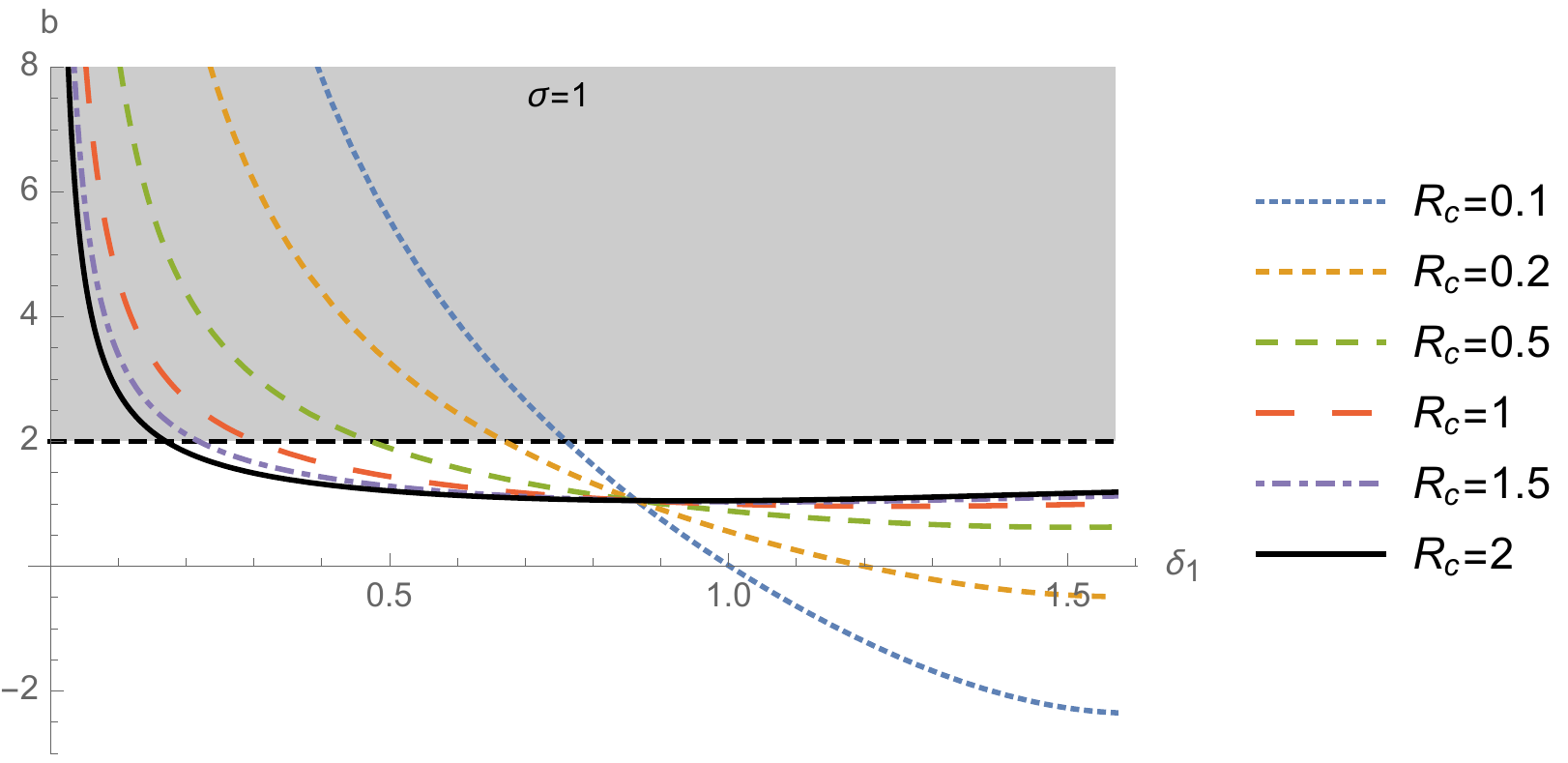}
    }\\
    \subfigure[]
    {
        \includegraphics[width=\columnwidth]{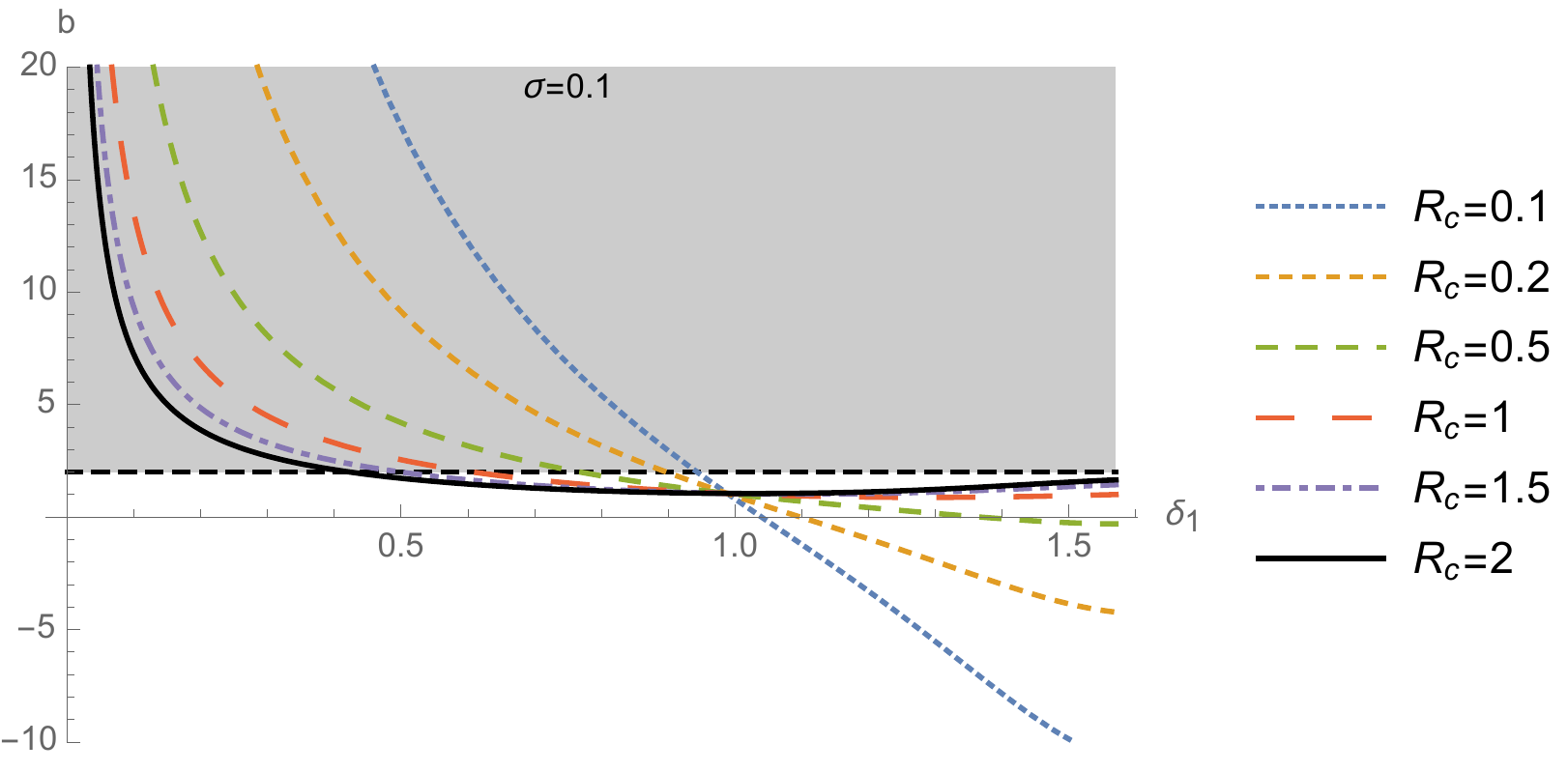}
    }
    \caption{The parameter $b\equiv a x_0/y_0$ as a function of the incident angle $\delta_1$ for different $\sigma$ and shock radius of curvature $R_c$. $R_c$ is normalized to $r_k$, namely the radius of the shock at the knot measured from the pulsar. Shaded regions correspond to physically favored value of $b$: $b>2$. (a) $\sigma=5$. (b) $\sigma=1$. (c) $\sigma=0.1$.}\label{fig:b}
\end{figure}

For straight outflow, there are several interesting points to make:
\begin{itemize}
    \item If we ignore the pitch angle dependence, namely, omit the $\sin^{\alpha}\varpi'$ factor in Equation (\ref{eq:emissivity}), from the Doppler factor we get the following:
    \begin{itemize}
        \item The width of the knot perpendicular to the symmetry axis is $\eta_{\perp}\sim 2r_k/\gamma_2$. For example, if $\alpha=0$ and $Y=0$, the integral can be carried out to give $I(\nu)=C_2(\nu)(1+\gamma_2^2X^2/x_0^2)^{-1}$. The intensity profile depends on $\alpha$: it is more peaked for a steeper spectral index, as shown in Figure \ref{fig:straightIprofile} (a).
        \item The ratio between the widths of the knot perpendicular and parallel to the symmetry axis $\eta_{\perp}/\eta_{\parallel}$ can be calculated as $a x_0/y_0=a\sin(\theta_{\rm ob}-\Delta_0)/[\sin\theta_{\rm ob}(\cos\Delta_0-\cot\delta_1\sin\Delta_0)]\equiv b$. Observationally we have $\eta_{\perp}/\eta_{\parallel}\approx2$. A contour plot of the knot intensity map is shown in Figure \ref{fig:straightI} (a) for the case $b=2$. The parameter $b$ depends on the shock radius of curvature $R_c$, incident angle $\delta_1$ and $\sigma$. Figure \ref{fig:b} shows $b$ as a function of $\delta_1$ for different $\sigma$ and $R_c$. For the case of quasi-parallel incident ($\delta_1\to0$), $b\sim a/\chi$ and the requirement on $R_c$ becomes the following if we want $\eta_{\perp}/\eta_{\parallel}\approx2$
            \begin{equation}\label{eq:Rc_parallel}
                R_c\sim\frac{r_k(1-\chi)}{(b-1)\chi\sin\delta_1}\sim\frac{r_k(1-\chi)}{\chi\sin\delta_1}.
            \end{equation}
            The trend is that for very high $\sigma$ flows, one needs very small $\delta_1$ and/or $R_c$ to get large enough b. This seems to be another difficulty associated with a very high $\sigma$ model.
    \end{itemize}
    \item When we take into account the pitch angle dependence of synchrotron power, $\eta_{\perp}$ decreases because the synchrotron emissivity depends upon the perpendicular component of the magnetic field. This effect is shown in Figure \ref{fig:straightIprofile} (b) and Figure \ref{fig:straightI} (b). For a spectral index $\alpha=0.8-1$, $\eta_{\perp}$ can become as small as 0.4 times the value estimated in \S\ref{sec:size} and this helps to reduce $\eta_\perp/\Psi_p$. However, $\eta_{\parallel}$ is not reduced by this effect; in order to still get the right aspect ratio $\eta_{\perp}/\eta_{\parallel}$, we would require $b$ to be 2-3 factors larger than that given in the last paragraph.
\end{itemize}

\subsubsection{Curved outflow}
\begin{figure*}
    \centering
    \subfigure[]
    {
        \includegraphics[width=0.3\textwidth]{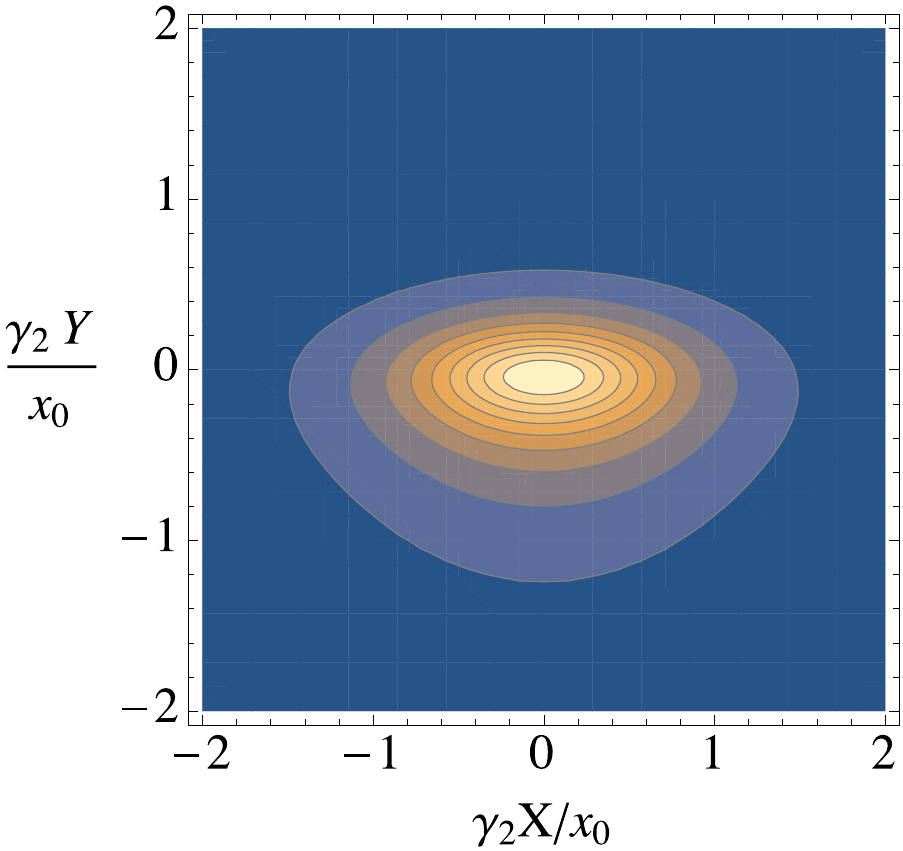}
    }
    \subfigure[]
    {
        \includegraphics[width=0.3\textwidth]{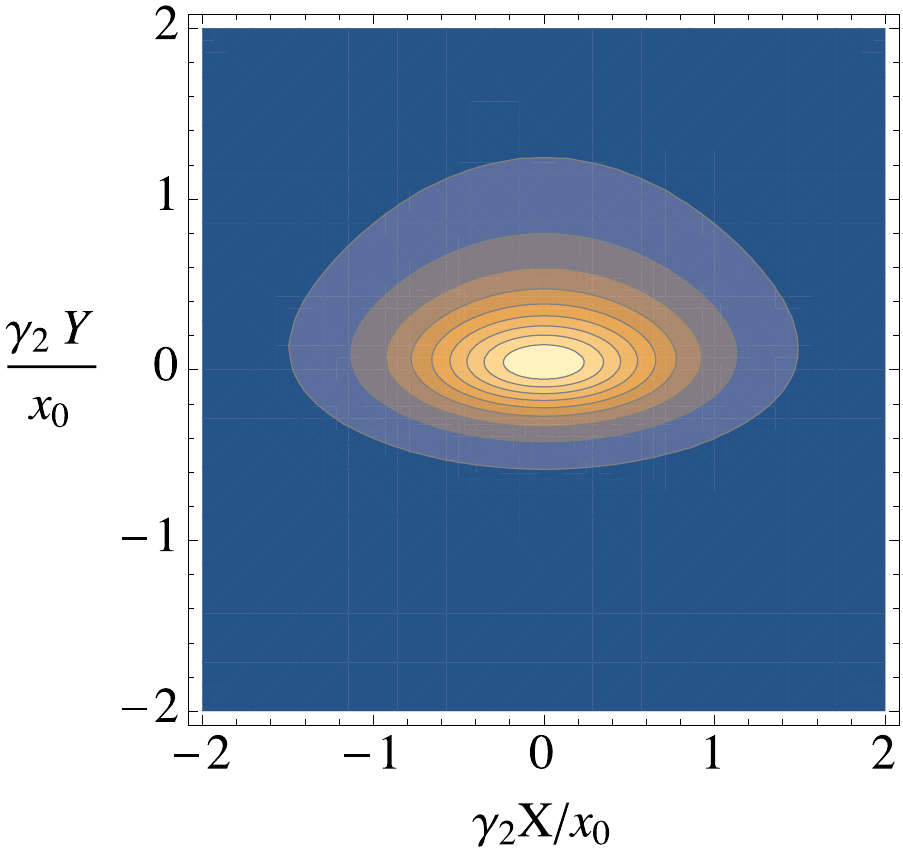}
    }
    \subfigure[]
    {
        \includegraphics[width=0.3\textwidth]{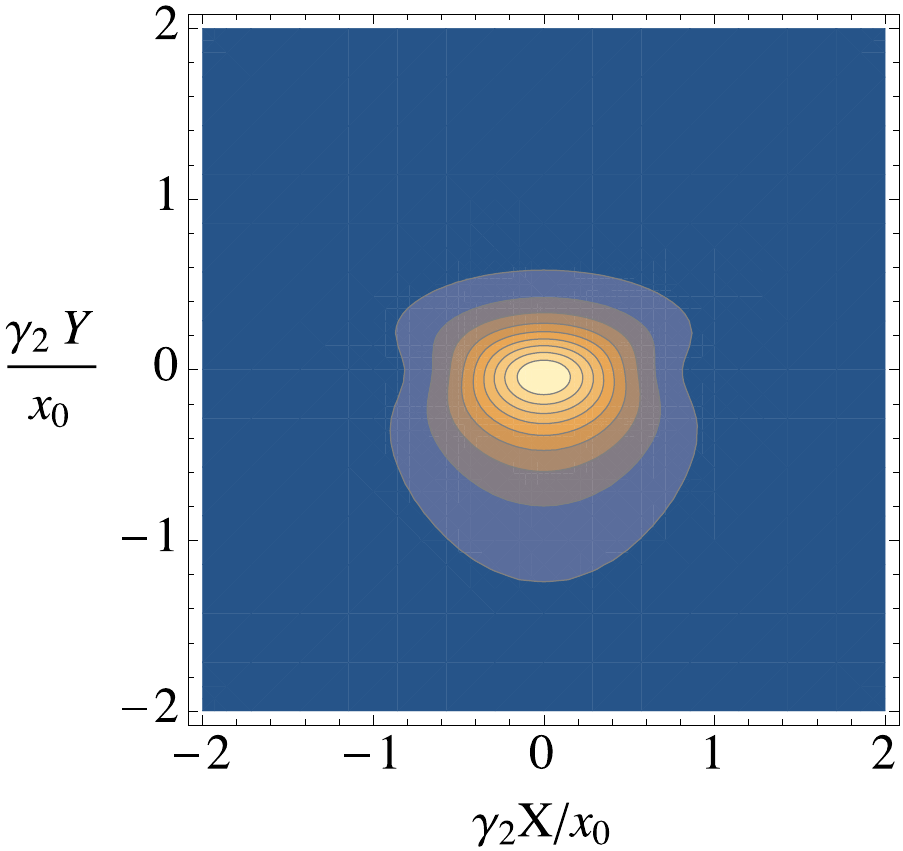}
    }
    \caption{Knot shape and intensity profile for the case of curved outflow, in the coordinate system of Figure \ref{fig:projection_coordinates}. The pulsar should be located somewhere on the positive y axis. (a) Contour Plot for the case $b=3$, $\alpha=0.8$, $\gamma_2\tau=-1/2$ (namely, the flow is bending toward the equator), when the pitch angle dependence is neglected. (b) Same as (a), except that $\gamma_2\tau=1/2$ (namely, the flow is bending toward the pole). (c) Same as (a), but with the pitch angle dependence of synchrotron power included.}\label{fig:curvedI}
\end{figure*}
For a curved outflow as introduced in Equation (\ref{eq:bending}), a few examples are shown in Figure \ref{fig:curvedI}. It can be seen that when the flow bends toward the equator, the knot is more extended on the side away from the pulsar; in contrast, if the flow bends toward the pole, it is the side close to the pulsar that becomes more extended. The effect of pitch angle dependence is still to contract the emission in the x-direction.

This effect may be important in explaining the curvature of the image shown in Keck and Hubble observations. If we imagine the outflow could bend through an angle $1/\gamma_2$ over a distance $\lesssim r_k$, Figure \ref{fig:curvedI} (a) shows that bending toward the equator could have the potential to give the right curvature in the image. The effect can be tested using more realistic models of the post-shock flow.

\subsection{Anisotropic particle distribution}\label{subsec:aniso}
Imagine that downstream of the shock the particle distribution is anisotropic and follows double adiabatic evolution, then the intensity profile would be different from the isotropic case above. One complication here is that, as the plasma element moves along the flow line, the pitch angle distribution also evolves. For simplicity, assume that the particles conserve their two adiabatic invariants so that the perpendicular and parallel momentum satisfy $p_{\perp}\propto A^{-1/2}\propto B^{1/2}$ and $p_{\parallel}\propto L^{-1}\propto nB^{-1}$. The first relation is easily justified as the particle Larmor radius is much smaller than the length scale of the shock, while the second requires sufficiently low collision rate as particles move along the toroidal magnetic field. When this is the case, the parallel and perpendicular pressure satisfy $d/d\tau (P_{\parallel}Bn^{-2})=0$ and $d/d\tau (P_{\perp}n^{-1}B^{-1/2})=0$. If as before $B\propto r^{-1}$ and $n\propto r^{-2}$, we would have $P_{\parallel}\propto r^{-3}$ and $P_{\perp}\propto r^{-5/2}$---$P_{\parallel}$ decreases faster than $P_{\perp}$. A proper treatment should evolve the distribution function along the flow line. Here, for simplicity and illustration, we assume that the emission is dominated by a thin layer right at the shock so it is instructive to look at the emissivity at the shock front.

\begin{figure*}
    \centering
    \subfigure[]
    {
        \includegraphics[width=0.3\textwidth]{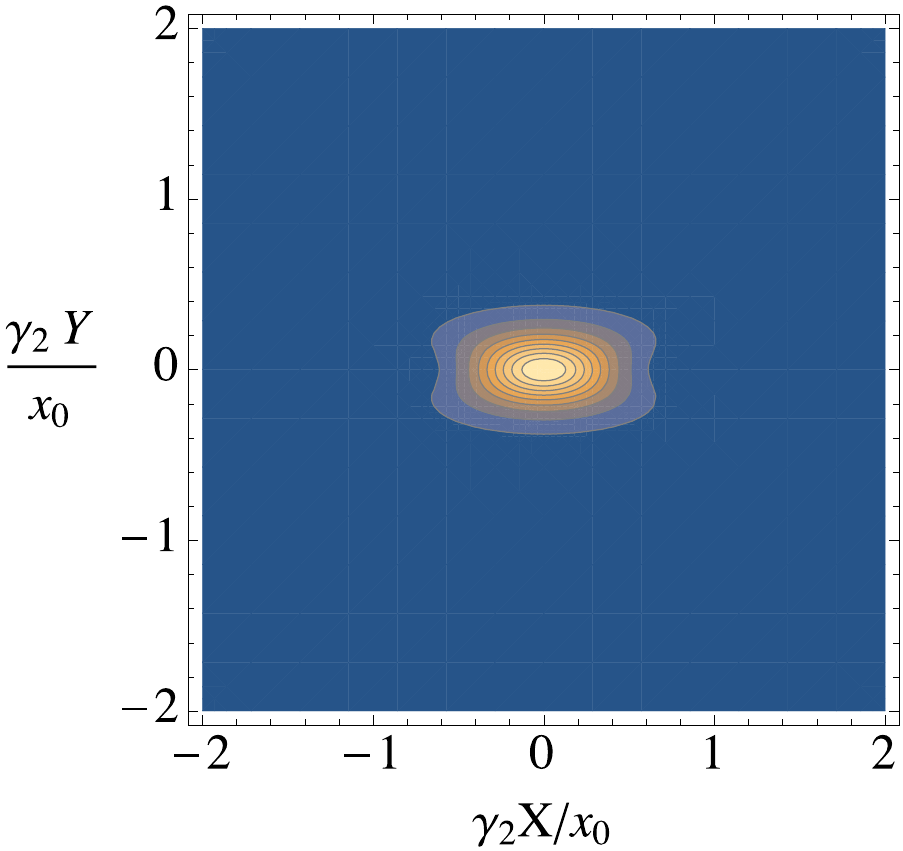}
    }
    \subfigure[]
    {
        \includegraphics[width=0.3\textwidth]{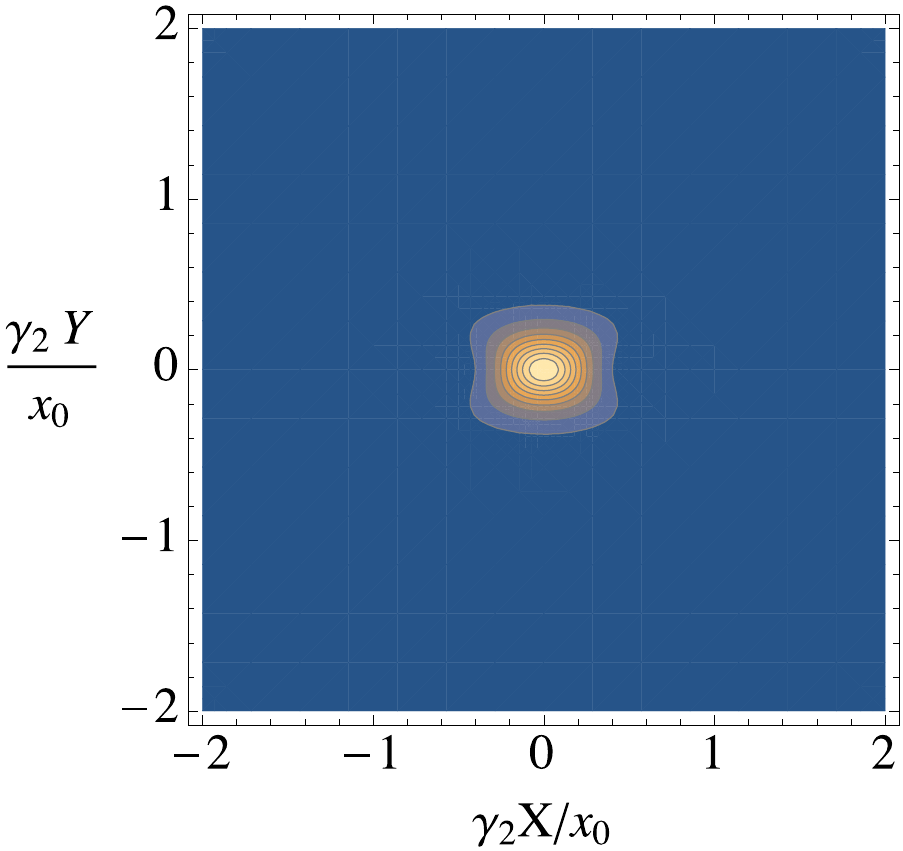}
    }
    \subfigure[]
    {
        \includegraphics[width=0.3\textwidth]{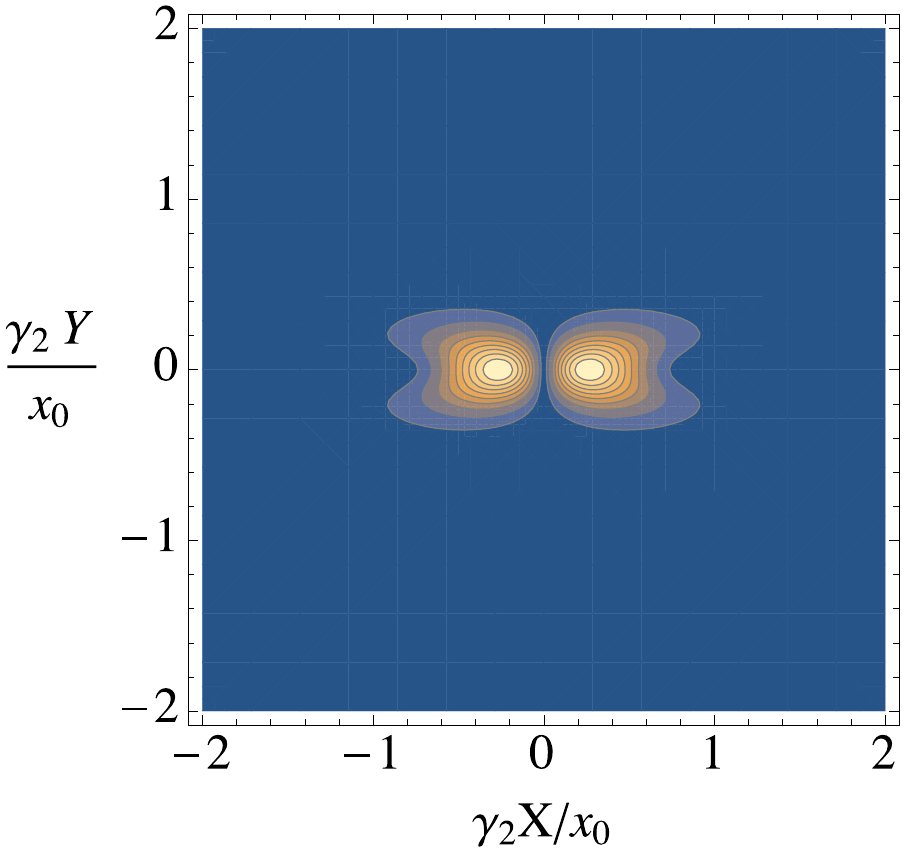}
    }
    \caption{Contour plot of the emissivity on the shock surface, as projected on the plane of the sky, for different functional forms of particle pitch angle distribution, according to Equation (\ref{eq:emissivity_shock_aniso_approx}). In these plots we take $b=3$, $\alpha=0.8$.  (a) Isotropic case, namely, $g=1$. (b) $g(\sin\varpi')=\sin^q\varpi'$ with $q=4$. (c) $g(\sin\varpi')=(1-\sin^2\varpi')^{q/2}$ with $q=1$.}\label{fig:j_aniso}
\end{figure*}

Suppose the particle distribution function can be written as 
\begin{equation}\label{eq:n_aniso}
f'(\gamma',\varpi')d\gamma'd\Omega'dV'=k'\gamma'^{-(2\alpha+1)}g(\sin \varpi')d\gamma'd\Omega'dV',
\end{equation}
where $g(\sin\varpi')$ describes the pitch angle distribution. We have the emissivity
\begin{equation}\label{eq:emissivity_aniso}
\begin{split}
    j_{\nu,\Omega}(\nu,\mathbf{n})&= \mathscr{D}^{2+\alpha}j'_{\nu',\Omega'}(\nu,\mathbf{n}')\\
    &=C(\nu) k'\mathscr{D}^{2+\alpha}(B'\sin\varpi')^{1+\alpha}g(\sin \varpi').
\end{split}
\end{equation}
When the downstream Lorentz factor $\gamma_2$ is large enough, the emissivity at the shock front can be approximated as
\begin{multline}\label{eq:emissivity_shock_aniso_approx}
j\approx C(\nu) k' B'^{1+\alpha} \left[\frac{2\gamma_2}{1+\gamma_2^2\psi^2}\right]^{2+\alpha} \left[1-\frac{4\sin^2\theta_{\rm ob}\gamma_2^2\phi^2}{(1+\gamma_2^2\psi^2)^2}\right]^{(1+\alpha)/2}\\
\times g\left(\left[1-\frac{4\sin^2\theta_{\rm ob}\gamma_2^2\phi^2}{(1+\gamma_2^2\psi^2)^2}\right]^{1/2}\right)
\end{multline}
We show a few contour plots of the emissivity for different functional forms $g(\sin\varpi')$ in Figure \ref{fig:j_aniso}. It can be easily seen that when $P_{\perp}$ dominates, the emission is more squeezed in the x direction and vice versa. (The parallel pressure dominant case may produce bipolar features.) If we focus on the former and assume $g(\sin\varpi')=\sin^{q}\varpi'$, the width of the knot perpendicular to the symmetry axis can now be estimated as $\eta_{\perp}\sim2r_k/\gamma_2/\sqrt{4+3\alpha+2q}$ according to Equation (\ref{eq:emissivity_shock_aniso_approx}). In order to obtain $\eta_{\perp}/\Psi_p\approx0.5$, we need (1) $q\sim15$, namely, the pitch angle distribution is highly concentrated near 90$^{\circ}$, (2) $\sigma\sim0$ and (3) $\delta_1\sim0$. At the same time, to keep the right aspect ratio $\eta_{\perp}/\eta_{\parallel}\approx2$, the radius of curvature of the shock surface should satisfy $R_c\lesssim0.2r_k/\sin\delta_1$, while $r_k=\Psi_p D/\Delta$ should be less than $r_{eq}$. These strong requirements make the knot to be a very special point on the shock surface, and it's doubtful whether such conditions can be always satisfied as the shock moves around due to the change of nebula pressure. The large perpendicular pressure is also not generic when there's sufficient scattering in the flow. This remains hard to understand.

\subsection{Polarization}\label{subsec:polarization}

In this subsection we examine in detail the Doppler-depolarization effect and suggest possible solutions. Let's consider a plasma element with spherical coordinates $\{r(\theta),\theta,\phi\}$ at the shock front. We follow the notations of Section \ref{subsec:emissivity@shock}. The emission has linear polarization perpendicular to the magnetic field in the comoving frame of the plasma, namely, $\hat{e'}={\bf n'}\times\hat{B'}$. Upon a Lorentz transformation back to the nebula frame, the direction of the electric field becomes \citep{Lyutikov:2003aa}
\begin{equation}\label{eq:E_direction}
    \hat{e}=\frac{{\bf n}\times(\hat{B}+{\bf n}\times(\vec{\beta_2}\times\hat{B}))}{\sqrt{(1-{\bf n}\cdot\vec{\beta_2})^2-(\hat{B}\cdot{\bf n})^2/\gamma_2^2}}
\end{equation}
since $\hat{B}\cdot\vec{\beta_2}=0$ in our case. Introduce $\hat{l}=\hat{y}$ which is a constant unit vector perpendicular to the plane containing ${\bf n}$ and $\hat{z}$, the polarization position angle $\tilde{\chi}$ is determined by

\begin{align}\label{eq:position angle}
    \cos\tilde{\chi} &= \hat{e}\cdot({\bf n}\times\hat{l})\nonumber\\
     &= \frac{\cos{\phi}(1-\beta_2\cos{\psi})-\beta_2\sin\theta_2\sin\theta_{\rm ob}\sin^2\phi}{\sqrt{(1-\beta_2\cos{\psi})^2-\sin^2\theta_{\rm ob}\sin^2\phi/\gamma_2^2}},\\
    \sin\tilde{\chi} &= \hat{e}\cdot\hat{l}\nonumber\\
    &= \frac{\sin{\phi}(\beta_2\cos\theta_2-\cos\theta_{\rm ob})}{\sqrt{(1-\beta_2\cos{\psi})^2-\sin^2\theta_{\rm ob}\sin^2\phi/\gamma_2^2}}.
\end{align}

Still assume that particle distribution satisfies Equation (\ref{eq:distribution}) and let $p=2\alpha+1$ in the following. We consider steady state and suppose that radiation is mostly emitted by a thin shell with thickness $\delta r$ right after the shock. The latter assumption is justified if the flow does not slow down too rapidly and roughly follows spherical expansion, as shown in \S\ref{subsec:flowdynamics}. Now we can calculate the Stokes parameters by integrating over the visible surface of the shock front:
\begin{multline}\label{eq:I}
I = \frac{p+7/3}{p+1}A(\nu)\frac{\delta r}{D^2}\int_0^{2\pi}d\phi\int_0^{\pi}d\theta \sin\theta \,r\sqrt{r^2+(dr/d\theta)^2}\\
\times\mathscr{D}^{2+(p-1)/2}k'|B'\sin\varpi'|^{(p+1)/2},
\end{multline}
\begin{multline}\label{eq:Q}
Q = A(\nu)\frac{\delta r}{D^2}\int_0^{2\pi}d\phi\int_0^{\pi}d\theta \sin\theta \,r\sqrt{r^2+(dr/d\theta)^2}\\
\times\mathscr{D}^{2+(p-1)/2}k'|B'\sin\varpi'|^{(p+1)/2}\cos2\tilde{\chi},
\end{multline}
\begin{multline}
U = A(\nu)\frac{\delta r}{D^2}\int_0^{2\pi}d\phi\int_0^{\pi}d\theta \sin\theta \,r\sqrt{r^2+(dr/d\theta)^2}\\
\times\mathscr{D}^{2+(p-1)/2}k'|B'\sin\varpi'|^{(p+1)/2}\sin2\tilde{\chi},
\end{multline}
\begin{equation}
V = 0,
\end{equation}
where
$$A(\nu)=\frac{\sqrt{3}}{4}\,\Gamma\!\left(\frac{3p-1}{12}\right)\Gamma\!\left(\frac{3p+7}{12}\right)\frac{e^3}{mc^2} \left(\frac{2\pi mc}{3e}\right)^{-(p-1)/2}\nu^{-(p-1)/2}.$$

Since U integrates out to 0 under axisymmetry, the polarization degree is $\Pi=Q/I$. We can get some analytical results under ultrarelativistic limit. In this case due to the Doppler factor $\mathscr{D}$, only a small region $\phi\sim1/\gamma_2$, $\theta_2-\theta_{\rm ob}\sim1/\gamma_2$ contributes significantly to the integral, so we can allow the integral limits to extend from $-\infty$ to $+\infty$. Also assume $\gamma_2$ itself as well as other quantities, $B_1$ and $n_1u_1$, do not change significantly over this small patch. We change the variables to $\xi=\gamma_2\phi$ and $\eta=\gamma_2(\theta_2-\theta_{\rm ob})$, carrying out the expansion in the limit $1/\gamma_2\ll1$, $\xi\sim O(1)$, $\eta\sim O(1)$, we have
\begin{gather}
\mathscr{D}\approx\frac{2\gamma_2}{1+\eta^2+\xi^2\sin^2\theta_{\rm ob}},\\
\sin\varpi'\approx\left[1-\frac{4\xi^2\sin^2\theta_{\rm ob}}{(1+\eta^2+\xi^2\sin^2\theta_{\rm ob})^2}\right]^{1/2},\\
\cos2\tilde{\chi}=\frac{(1+\eta^2-\xi^2\sin^2\theta_{\rm ob})^2-4\eta^2\xi^2\sin^2\theta_{\rm ob}}{(1+\eta^2+\xi^2\sin^2\theta_{\rm ob})^2-4\xi^2\sin^2\theta_{\rm ob}}.
\end{gather}
Also note that $d\theta_2=a\,d\theta$ where $a=\chi\cos^2\delta_2/\cos^2\delta_1+(1-\chi\cos^2\delta_2/\cos^2\delta_1)r_k/(\sin\delta_1 R_c)$, so the integral over $\theta$ can be written into an integral over $\theta_2$. 
Let $\xi_1=\xi\sin\theta_{\rm ob}$. When $\theta_{\rm ob}\gg1/\gamma_2$, the result of polarization degree no longer depends on $\theta_{\rm ob}$:
\begin{multline}\label{eq:PD}
    \Pi=\frac{Q}{I}=\frac{p+1}{p+7/3}
    \Bigg\{\int_{-\infty}^{+\infty}\!d\xi_1\int_{-\infty}^{+\infty}\!d\eta (1+\eta^2+\xi_1^2)^{-(2+p)}\\
    \times[(1+\eta^2+\xi_1^2)^2-4\xi_1^2]^{(p-3)/4}[(1+\eta^2-\xi_1^2)^2-4\eta^2\xi_1^2]\Bigg\}\\
    \div \int_{-\infty}^{+\infty}\!d\xi_1\int_{-\infty}^{+\infty}\!d\eta (1+\eta^2+\xi_1^2)^{-(2+p)}[(1+\eta^2+\xi_1^2)^2-4\xi_1^2]^{(p+1)/4}
\end{multline}
This gives the same result as \citet{Lyutikov:2003aa}. When $p=3$, $\Pi=9/16=56.25\%$, and when $p=2$, $\Pi=43.4\%$. For the oblique termination shock of the Crab Nebula, we expect the particle distribution power law index $p$ to be between 2 and 3, thus the theoretical upper limit for the polarization degree, under ultrarelativistic assumption, would be $\lesssim56\%$. In particular, for the best-guess spectral index $\alpha=0.8$, $p=2.6$, we have $\Pi=51.6\%$. 

However, the flow may only be mildly relativistic, so the aberration would also be modest. In this case a more careful treatment is necessary and the result should depend on the details of the shock geometry. As an example we consider the fiducial shock shape $r=\sin^4\theta$ as shown in Figure \ref{fig:emissivity_shock}, where the downstream flow is only mildly relativistic, with Lorentz factor $\gamma_2\sim3-5$ at the knot (see Figure \ref{fig:emissivity_shock}a). We evaluate the Stokes parameters $I$ and $Q$ in Equations (\ref{eq:I})(\ref{eq:Q}) by integrating the emissivity $j$ over the surface area where $j\ge j_{\nu,\rm{peak}}/10$: we caution that this size is much larger than the observed value (see Figure \ref{fig:emissivity_shock}b), as expected. Nevertheless, we proceed with the polarization calculation. We find that when $p=3$, $\Pi=68.9\%$; when $p=2$, $\Pi=61.2\%$; when $p=2.6$, $\Pi=66.2\%$. The results seem to be consistent with the observed polarization degree, but as we already noted, this is an artificial example that does not give the right compactness and curvature. To solve the polarization problem we need to understand the other two puzzles simultaneously. 

Clearly more careful modeling of the post-shock flow and its emission is needed. As the polarization degree is sensitive to the spectral index and the flow Lorentz factor, we could in principle constrain them from observed polarization degree. Observationally, it would be very helpful to get an accurate, simultaneous measurement of the spectral index.

\section{Fluxes of the knot in different wave bands}\label{sec:flux}
Having explored the challenges to the shock model in terms of the compactness, curvature and polarization of the knot, we now turn to other signature properties of the knot that can equally set important constraints on general shock models. The first one among these is the flux of the knot in different wavebands, which indicates the amount of dissipation in the relativistic outflow. It is commonly presumed that the termination shock is responsible for accelerating most of the IR- to X-ray-emitting particles, but it's not yet well understood what mechanism could act efficiently enough at such a relativistic, perpendicular shock. Another very interesting question is whether the shock is the sole source for accelerating IR/optical emitting particles, or some in situ acceleration further out in the body of the nebula is needed. In this section we try to get some hints to these questions by characterizing the radiation efficiency of the knot and deducing the energy composition of the shock outflow.

\subsection{Observations}
Currently we have flux data in IR, optical and upper limits in radio, $\gamma$-ray bands:

\begin{itemize}
  \item IR (Keck): \citet{Weisskopf:2013ab} give a typical $K'$ band knot magnitude $K'=15.9$, after dereddening with the best fit value $E(B-V)=0.52$ \citep{Sollerman:2000aa} and the extinction law from \citet{Fitzpatrick:1999aa}, this corresponds to $F_{\nu}^{\text{knot}}=3.1\times 10^{-27}\text{erg}\, \text{s}^{-1} \text{cm}^{-2} \text{Hz}^{-1}$. We define $\epsilon(\nu)\equiv\nu L_{\nu,\Omega}/(\dot{E}/4\pi)$, where $\dot{E}=5\times10^{38}\ {\rm erg}\,{\rm s}^{-1}$ is the pulsar spin down power and $L_{\nu,\Omega}=D^2F_{\nu}^{\text{knot}}$ is the knot luminosity per steradian, as the radiation efficiency of the shock along the line of sight in the particular waveband, and so $\epsilon (1.4\times10^{14}\,\rm{Hz})=4.2\times10^{-7}$.

      The latest Keck AO observations by \citet{Rudy:2015aa} give the knot-pulsar flux ratio $\sim0.063$ on average. If we take the photometry of the pulsar from \citet{Sandberg:2009aa}, the pulsar $K_s$ band magnitude is $13.80\pm0.01$, this corresponds to a dereddened flux of $2.2\times 10^{-26}\text{erg}\, \text{s}^{-1}\text{cm}^{-2}\text{Hz}^{-1}$, giving a knot flux in the $K'$ band $F_{\nu}^{\text{knot}}=1.4\times 10^{-27}\text{erg}\, \text{s}^{-1}\text{cm}^{-2}\text{Hz}^{-1}$.
  \item Optical (Hubble): The latest analyses of HST monitoring of the Crab \citep{Rudy:2015aa} give the dereddened green knot flux on average $1.0\times 10^{-27}\text{erg}\, \text{s}^{-1}\text{cm}^{-2}\text{Hz}^{-1}$, implying a radiation efficiency of $\epsilon (5.4\times10^{14}\,\rm{Hz})=5\times10^{-7}$, similar to the infrared estimate.
  \item Radio (VLA): The latest VLA observation \citep{Bietenholz:2015aa} set an upper limit $F_{\nu}^{\text{knot}}< 300 \mu\text{Jy}$ over a beam size of $1\arcsec\times1\arcsec$. Thus the radiation efficiency is $\epsilon (5.5\times10^{9}\,\rm{Hz})<1.6\times10^{-11}$.
  \item $\gamma$-ray (Fermi): Although there isn't enough spatial resolution to detect structures within the nebula, the total synchrotron $\gamma$-ray flux of the whole nebula in quiescent state gives an upper limit on any possible synchrotron $\gamma$-ray emission from the knot. According to \citet{Buehler:2012aa}, the quiescent nebula synchrotron emission has an integral photon flux above 100 MeV of  $(6.1\pm0.2)\times10^{-7}\, \rm{cm}^{-2}\, \rm{s}^{-1}$ and a photon index of  $3.59 \pm 0.07$. This corresponds to an energy flux above 100 MeV $F(\nu>2.4\times10^{22}\,\rm{Hz})=1.6\times 10^{-10}\,\rm{erg}\,\rm{cm}^{-2}\,\rm{s}^{-1}$. So the upper limit on the radiation efficiency of the knot in $\gamma$-rays is $\epsilon(\nu>2.4\times10^{22}\,\rm{Hz})<1.2\times10^{-5}$.
\end{itemize}

\subsection{Emissivity and particle enthalpy fraction}
For a broad spectrum of electrons/positrons, the pressure will be dominated by the part of the distribution that has a power law index of 2, which corresponds to an emission spectrum $F_{\nu}\propto\nu^{-0.5}$ locally. A few spectral measurements have been done so far, for example, \citet{Sollerman:2003aa} and \citet{Melatos:2005aa} give an index 0.8 from IR to NUV. Some more recent ones show slightly different spectral index ranging from 0.63 \citep{Tziamtzis:2009aa} to 1.3 \citep{Sandberg:2009aa} but unfortunately these are based on non-simultaneous data. The spectrum of the nebula has an index $\sim0.6-0.8$ in the same band. If the knot index is indeed of similar value, the IR/optical emitting particles should contribute a major fraction of particle pressure in the shock outflow.

We estimate the particle enthalpy fraction in relevant energy bands from the radiation efficiency derived above. Suppose that the emitting area is $A\sim(r_k/\gamma_2)^2$, and the emissivity drops off significantly over a distance $r_k/\varsigma$, then we have
\begin{equation}
\nu L_{\nu,\Omega}\sim\nu j_{\nu,\Omega}Ar_k/\varsigma=\left(\frac{\nu}{\nu'}\right)^3\nu'j_{\nu',\Omega'}'Ar_k/\varsigma\sim\gamma_2\nu'j_{\nu',\Omega'}'r_k^3/\varsigma.
\end{equation}
Assuming an isotropic particle distribution in the fluid rest frame, we obtain
\begin{equation}
\nu'j_{\nu',\Omega'}'\sim \frac{1}{2}\,n\,2\sigma_Tc\gamma'^2\,\frac{B'^2}{8\pi}=\frac{n\gamma'mc^2}{2t'_{\text{cool}}},
\end{equation}
where $t'_{\text{cool}}=6\pi  \epsilon _0m^3c^3/(e^4B'^2\gamma ')$ is the synchrotron cooling time of the particles in the fluid rest frame. Now the enthalpy fraction of the particles that emit at a particular photon frequency $\nu$ is
\begin{equation}
    \eta(\gamma')=\frac{\frac{4}{3}n\gamma'mc^2\gamma_2^2r_k^2c}{\dot{E}/(4\pi)}.
\end{equation}
Using the above results we get
\begin{equation}
\eta(\gamma')=\frac{8}{3}\,\varsigma\,\frac{\gamma_2 t'_{\text{cool}}}{r_k/c}\,\frac{\nu  L_{\nu ,\Omega }}{\dot{E}/(4\pi)}=\frac{8}{3}\,\varsigma\,\frac{\gamma_2 t'_{\text{cool}}}{r_k/c}\,\epsilon.
\end{equation}
From this we can also estimate the particle injection rate per unit steradian
\begin{equation}
\dot{N}(\gamma')=\frac{3\eta(\gamma') }{4 \gamma ' m c^2}\frac{\dot{E}}{4\pi\gamma _2}.
\end{equation}
Taking a typical downstream magnetic field $B_2=B_{-3}$ mG, $r_k=r_{17}$ cm, Lorentz factor $\gamma_2\sim5$, and pulsar spin down luminosity $\dot{E}=5\times10^{38}\;\rm{erg}\;\rm{s}^{-1}$, we obtain the particle injection rate and enthalpy fraction in different energy range as follows:
\begin{itemize}
  \item IR-emitting particles:
  \begin{align}
  \dot{N}_{\rm{IR}}&=4.6\times10^{36}\;\text{s}^{-1}\text{sr}^{-1}\,\varsigma\, r_{17}^{-1}B_{-3}^{-1}(\gamma_2/5)^2,\\
  \eta_{\rm{IR}}&=0.11\,\varsigma (\gamma_2/5)^3B_{-3}^{-3/2} r_{17}^{-1};
  \end{align}
  \item Optical-emitting particles:
  \begin{align}
  \dot{N}_{O}&=1.5\times10^{36}\;\text{s}^{-1}\text{sr}^{-1}\,\varsigma\, r_{17}^{-1}B_{-3}^{-1}(\gamma_2/5)^2,\\
  \eta_{O}&=0.075\,\varsigma (\gamma_2/5)^3B_{-3}^{-3/2} r_{17}^{-1};
  \end{align}
  \item Radio-emitting particles:
  \begin{align}
  \dot{N}_{R}&<4.6\times10^{36}\;\text{s}^{-1}\text{sr}^{-1}\,\varsigma\, r_{17}^{-1}B_{-3}^{-1}(\gamma_2/5)^2,\\
  \eta_{R}&<7.2\times10^{-4}\,\varsigma (\gamma_2/5)^3B_{-3}^{-3/2} r_{17}^{-1}.
  \end{align}
   \item $\gamma$-ray emitting particles that give synchrotron photon at 100 MeV:
  \begin{align}
  \dot{N}_{\gamma}&<1.6\times10^{31}\;\text{s}^{-1}\text{sr}^{-1}\,\varsigma\, r_{17}^{-1}B_{-3}^{-1}(\gamma_2/5)^2,\\
  \eta_{\gamma}&<0.005\,\varsigma (\gamma_2/5)^3B_{-3}^{-3/2} r_{17}^{-1}.
  \end{align}
\end{itemize}
We know from observations that the total optical to X-ray emission of the whole Crab Nebula requires a particle injection rate of $10^{38.5}$ pairs/s and the radio emission requires an injection rate of $10^{41}$ pairs/s \citep[e.g.,][]{Arons:2012aa}. The above results indicate that the oblique part of the termination shock cannot provide enough radio-emitting particles. The number of IR/optical-emitting particles also seems insufficient. The IR emitting particle fraction $\eta_{\rm IR}\sim0.11$ is indeed substantial, indicating that they might be around the peak in the post shock particle distribution.

In order to investigate the acceleration mechanism, we can estimate how many orbits the particles complete to give the observed radiation power. Since the flow time scale is $t_{\rm flow}=r_k/(\varsigma c)$, and the gyro frequency of the particles is $\nu_g=\gamma_2\nu_g'=eB/(2\pi \gamma' m)$ in the nebula frame, we get the number of orbits to be (take IR band for example, and let $\nu=\nu_{14}10^{14}\,\rm{Hz}$)
\begin{equation}
\begin{split}
N&\sim\frac{r_k}{\varsigma c}\,\frac{eB}{2\pi m}\,\sqrt{\frac{3eB}{4\pi m\nu}}
\sim\frac{8}{3}\,\epsilon\,\frac{\gamma_2^3}{\eta}\,\frac{9c}{16\pi^2r_e\nu}\\
&=\frac{6\times10^4}{\varsigma}B_{-3}^{3/2} \nu_{14}^{-1/2}r_{17}.
\end{split}
\end{equation}
Therefore, IR/optical emitting particles could in principle be accelerated by mechanisms that operate over many gyration time scales. However, this would be difficult for $\gamma$-ray emitting particles.

\section{Variability}\label{sec:variability}

Another salient feature of the knot is its time variability. Both HST and Keck observations on timescales of months to years \citep{Rudy:2015aa} found that variations of the knot size $\eta_{\perp}$ and $\eta_{\parallel}$, surface brightness $I(\nu)$ and flux $F(\nu)$ correlate with the knot-pulsar separation $\Psi_P$. This is probably not that surprising as the inner nebula is known for its variable emission, especially the dynamic wisps, ever since the pioneering work of e.g. \citet{Scargle:1969aa}. Here we show that based on quite general grounds shock models can give a natural explanation to the variability of the knot.

In the shock model, variability can arise from at least two sources: changes in the surrounding nebula that result in modified boundary condition for the downstream flow, and variations in the upstream condition due to the pulsar. 

\subsection{Variations initiated downstream}\label{subsubsec:downstream_variation}

Inhomogeneous and unsteady pressure in the nebula can cause the shock radius and shape to vary over time. Numerical MHD simulations (e.g. 2D runs by \citealt{Camus:2009aa} and 3D runs by \citealt{Porth:2014aa}) show that even if the upstream condition is fixed, the post-shock flow is highly variable as observed, due to vortex shedding from the termination shock, and interaction with large amplitude waves in the nebula. The simulations of these complex variations exhibit timescales consistent with the observations.

Suppose that some change in the nebula pressure causes the shock radius $r_k$ and incident angle $\delta_1$ to change at the location of the knot. Assume the time scale to be long enough so that the shock evolves quasi-statically.  According to Equation (\ref{eq:emissivity}), $j(\nu)\propto k'\mathscr{D}^{2+\alpha}B'^{1+\alpha}$. Suppose $k'\propto n_2$, and since $n_2\sim n_1u_1/u_2\sim n_1u_1/\gamma_2$ (if $\gamma_2$ is large enough), $B'=B_2/\gamma_2=B_1/(\chi\gamma_2)$, at peak intensity point $\mathscr{D}\sim2\gamma_2$, we get $j\propto n_1u_1(B_1/\chi)^{1+\alpha}$. Also noticing that $n_1u_1\propto r_k^{-2}$, $B_1\propto r_k^{-1}$, we find $j\propto r_k^{-(3+\alpha)}$. Now suppose the emission length is $\ell$ (which may depend on $r_k$ too), then the peak intensity $I(\nu)\propto r_k^{-(3+\alpha)}\ell$. The knot-pulsar separation can be written as $\Psi_p=r_k\Delta/D$, so it's not surprising that $I(\nu)$ is correlated with $\Psi_p$. For example, if $\ell\propto r_k$ and if $\Delta$ weren't changed significantly, we would have $I(\nu)\propto \Psi_p^{-(2+\alpha)}$. Also the width $\eta_{\perp}\sim r_k/\gamma_2$, is again correlated with $r_k$ and thus $\Psi_p$, and roughly $\eta_{\perp}\propto \Psi_p$. Note that in addition to $r_k$, these properties also depend on $\gamma_2$, which is essentially changing with $\delta_1$ as can be seen from Equation (\ref{eq:gamma2}). The variation of the width parallel to the symmetry axis $\eta_{\parallel}$ is more involved as one needs to take into account the variation in the shock shape. \citet{Porth:2014aa} reported a correlation between knot flux and its separation from the pulsar that is similar to the observed one based on their 3D MHD simulation, suggesting that the shock model could give a good explanation of the knot variation.

\subsection{Variations initiated upstream}\label{subsubsec:upstream_variation}

Alternatively, a variation in the pulsar wind could cause variable knot emission. In such a scenario, the shortest possible variation time scale is $t_v\sim r_k(1-\cos (1/\gamma_2))/c\sim r_k/(2\gamma_2^2c)\sim 1\ \text{day} (r_k/10^{17}\,\text{cm})(\gamma_2/5)^{-2}$. Suppose the downstream pressure is fixed, while the upstream energy flux in the direction along the line of sight changes. In such a case, the shock radius and shape would still vary accordingly, but as $j(\nu)\propto k'\mathscr{D}^{2+\alpha}B'^{1+\alpha}\propto n_2\gamma_2B_2^{1+\alpha}$, it won't change much if $n_2\gamma_2$ and $B_2$ stay more or less the same. However, the knot-pulsar separation $\Psi_p=r_k\Delta/D$, and knot size $\eta_{\perp}\sim r_k/\gamma_2$ would depend on both $r_k$ and $\delta_1$; as a result, while the correlation between $\eta_{\perp}$ and $\Psi_p$ is similar to that in \S\ref{subsubsec:downstream_variation}, the correlation between $I(\nu)$ and $\Psi_p$ would be quite different.

\section{Discussions}\label{sec:discussion}
In this paper, we have presented a discussion of the implications of recent observations of the inner knot of the Crab Nebula for its interpretation as a flow immediately behind a pulsar wind termination shock \citep{Komissarov:2004aa, Komissarov:2011aa}. Some of these observations can be characterized as being broadly consistent with this interpretation; others present challenges to it. 

In the former category, we remark that some sort of feature is expected. Indeed, it is quite surprising that the radiative efficiency of the outflow of a supposedly ultrarelativistic wind in the spectral band where most of the pulsar power is radiated, at near UV frequency, is apparently so low ($\epsilon\lesssim10^{-6} $). (In principle, the wind could be more radiative in $\gamma$-rays, but the absence of a correlation of the knot variation with the $\gamma$-ray flares can be used to argue against this.) Another success of this model is that the knot is located on the projected nebular symmetry axis (presumably coincident with the pulsar spin axis), that it is elongated perpendicular to this axis and that the direction of the linear polarization, supposedly perpendicular to the expected toroidal magnetic field, is along the symmetry axis. In principle, these attributes could be associated with a mildly relativistic internal shock, or a Mach disk, in an inward extension of the southern jet, but it is then surprising that the observed dissipation is seen so close to the pulsar and that the possible other knot observed with this position angle relative to the pulsar is too far away ($\sim3.8\arcsec$, see \citealt{Hester:1995aa}). The variability on an outflow timescale is at least expected given the widespread, observed activity of the ``wisps'' in the inner nebula \citep{Scargle:1969aa,Hester:1995aa,Hester:2002aa}.  Furthermore the tentative correlation of the flux and location of the knot has a natural interpretation in terms of motion of the shock surface.

The first challenge to this interpretation is that the compactness of the knot compared to its offset from the pulsar ($\eta_{\perp}/\Psi_p\sim0.5$) and its elongation perpendicular to the symmetry axis ($\eta_{\perp}/\eta_{\parallel}\sim2$) were hard to reproduce quantitatively using the shock model. This may require the shock to have an almost grazing upstream velocity yet very small radius of curvature on the meridional crosssection ($R_c\lesssim 0.2r_k/\sin\delta_1$, see \S\ref{subsec:aniso}), which seems unnatural, or require the particle velocity distribution in the nebula frame to be elongated transversely, either through shear in the bulk velocity or a strong parallel pressure which is the opposite of what would be expected at a perpendicular shock front.  Even more perplexing is the curvature of the knot. It is concave towards, as opposed to away from, the pulsar. As we have described, it may be possible to account for this but only with a special flow field or observer orientation. Another great challenge is the degree of polarization. As we have explained even with a uniform toroidal field, the Doppler rotation effect will effectively reduce the synchrotron polarization of 0.73 to $\sim0.5$, significantly less than the reported value. 

However, the greatest challenge to this model may come from understanding how the particle acceleration actually takes place. In models of relativistic cosmic plasmas, it is common to presume that particle acceleration takes place at relativistic shocks in much the same way as it does at non-relativistic shocks with particles gaining energy as they are scattered back and forth across the shock front. However, the kinematics of relativistic shocks are fundamentally different and although transmitted distribution functions can be computed for given assumptions \citep[e.g.,][]{Gallant:1992aa,Hoshino:1992aa,Sironi:2009aa,Sironi:2011aa}, a fully consistent description of the acceleration eludes us.  There are two key questions that are raised by this interpretation. The first is why is the knot so strongly polarized if there are strong magnetic fluctuations close to the shock front and $\sigma<1$. Somehow these fluctuations have to damp out leaving behind a dynamically weak yet very uniform toroidal field. The second question is whether or not the shock is the dominant source of particles with energies above, say, 100 GeV emitting infrared synchrotron radiation or are most of the electrons and positrons accelerated much farther out in the nebula where most of the emission occurs. It is not possible to give a definitive answer to this question but we can estimate how fast the ultra-violet- \citep{Hennessy:1992aa} and X-ray- \citep{Weisskopf:2000aa,Seward:2006aa} emitting electrons would have to transport. If we suppose that the transport is essentially radially diffusive, then in order to reach the outer parts of the observed nebulae at these energies, the mean-free path perpendicular to the presumed toroidal field direction, must be $\sim10^2 (B/10^{-4}\, \rm{G})^3$  gyro radii (independent of emission frequency). This is sufficiently large and may suggest that some of the particle acceleration is actually {\it in situ}. 

One of the longest debates involving the Crab Nebula is the value of the magnetization parameter, $\sigma$.  There is little doubt that the torque applied to the neutron star is essentially electromagnetic and the outward energy and angular momentum flux in the vicinity of the light cylinder is mostly electromagnetic; otherwise it would be very difficult to account for the observed pulse profiles \citep[e.g.,][]{Arons:2009aa}. However the wind flows through nine decades of radius before its momentum flux matches the ambient nebula pressure and much can happen over this journey. Its initial electromagnetic structure is a combination of the axisymmetric, increasingly toroidal field associated with an aligned rotator and the striped, low latitude field reversing every $\sim5\times10^8$~cm of an oblique rotator. The wind radial velocity is expected to accelerate to ultrarelativistic velocity -- the velocity of the local Lorentz frame in which the local electric field vanishes. This can happen with minimal dissipation when the acceleration is due to the electromagnetic Lorentz force.  However, non-stationarity at the source and instability in the outflow will inevitably heat the gas -- we presume a pair plasma -- and limit the acceleration of the outflow. This is especially important at low latitude where the stripes may be able to reconnect with considerable dissipation \citep[e.g.,][]{Coroniti:1990aa} although strong arguments have been given that there is insufficient time in the comoving frame for this to be effective \citep[e.g.,][]{Lyubarsky:2001aa}. 

Consideration of the large scale current flow in the wind allows a complementary description to one based upon magnetic field. On quite general grounds, the unipolar induction mechanism will generate a potential difference $\sim50$~PV between the poles and the equator and there will be a quadrupolar poloidal current distribution, with a magnitude $\sim$500 TA in each hemisphere. The latitude-dependence of this current is what ultimately determines the shape of the shock surface. If the current ``closes''  well within the shock, the magnetization will be low and {\it vice versa}. When $\sigma$ is low, the ratio of the bulk acceleration to the ``Ohmic'' dissipation dictates the Mach number and asymptotic fluid Lorentz factor. Unfortunately we do not have a good enough understanding of these subtle issues to predict $\sigma$ and $\gamma$ and must rely on observational guidance.

We also have some clues from the nebula beyond the shock. What we observe is that electrons and positrons are accelerated to TeV energies and these particles ultimately radiate most of the rotational energy extracted from the neutron star. They do so on time scales $\sim30$~ yr., long compared with the flow time from the pulsar but slow compared with the lifetime of the nebula with high average efficiency although we know that energy can build up in the nebula and then be released faster than it is supplied \citep{Wilson-Hodge:2011aa}. There is, of course, a broad electron distribution from radio-emitting GeV electrons to the $\sim3$~ PeV $\gamma$-ray synchrotron-radiating electrons that produce the dramatic flares. (These also radiate $\sim0.01$ of their power as inverse Compton $\gamma$-rays.) The low energy electrons accumulated over the nebula lifetime while the high energy electrons have to be accelerated locally as their cooling times are so short. At the highest energies they cool in a small fraction of a Larmor orbit. On these grounds, it has been argued that some of the nebula must be magnetically-dominated as this seems to be necessary to account for such dramatic, impulsive acceleration. However, it has also been argued \citep{Atoyan:1996aa,Meyer:2010aa}, on the basis of the ratio of the Compton to the synchrotron power, that, on average, the particle energy density, dominated by TeV electrons, exceeds the magnetic energy density.  This suggests that the nebula is quite inhomogeneous. The relatively high average nebula polarization in the optical and X-ray bands ($\sim0.2$, see, e.g. \citealt{Woltjer:1957aa,Oort:1956aa,Schmidt:1979aa,Weisskopf:1978aa}) indicates a long range order again suggesting that $\beta\equiv8\pi P/B^2$ is not much greater than unity.

Returning to the physical conditions at the inner knot, interpreted as a shock, we have argued that the compactness combined with the shape implies a low magnetization, $\sigma<1$. One way to reconcile this with the indication that some of the nebula is highly magnetized is if the line of sight passes through the striped wind at small radius and the dissipation of this magnetic energy is efficient (e.g. \citealt{Komissarov:2013aa,Porth:2013aa}). However, this seems inconsistent with the high polarization. We therefore consider some alternative possibilities. The first of these is that the post shock flow is more complicated and typically contains a point of inflexion after it is pulled towards the poles by the magnetic tension. The path length along the flow will be long at these inflexions and the single flow line when the tangent is also directed towards us is the location of the knot. This can be examined using numerical simulations of the post-shock flow. The second idea is that the pre-shock flow is non-radial due to the action of magnetic stress and the knot is formed where a special path perhaps associated with a current sheet in the magnetosphere becomes tangent to the line of sight.

The third possibility is more radical, that the outflow is highly magnetized and there is no shock. When most of the pressure and current is associated with the highest energy particles that are radiating efficiently, a simple fluid description is called into question. The flow is better described using individual particle orbits and the deceleration is gradual. Well beyond the light cylinder the magnetic field is essentially toroidal and the poloidal current is composed of electrons and positrons undergoing gradient, curvature and magnetization drifts. It is possible that the details of these motions are affected by radiation reaction operating on the highest energy particles. As the guiding centers of the orbits move along the electric field, the particles gain energy and a self-consistent description of the particle acceleration and emission becomes possible. The nature of the inner knot is then associated with the non-radial motion of the guiding centers that leads to a maximum intensity on the sky. This basic idea can be explored using PIC codes.

We note that \citet{Lyutikov:2015aa} have carried out a similar investigation of the knot emission as coming from the termination shock, and reached the conclusion that an effectively low $\sigma$ shock can produce all the observed properties of the knot. They use the Kompaneets approximation where it is presumed that the post-shock flow is isobaric to model the shape of the shock; this only explores a quite limited parameter space as compared to our more general analytical expansion around the peak, while the peak intensity may be close to the boundary of the shock so that adding up the emissivity of the two folded sections of the shock may help produce a small enough $\eta_{\parallel}$. However, the size perpendicular to the symmetry axis, $\eta_{\perp}$, still appears larger than the observed value in their model, and the curvature of the knot is inconsistent with current observation. \citet{Lyutikov:2015aa} argue that finite thickness of emitting region and/or limited resolution of the observation could change the curvature. They also point out that \citet{Moran:2013aa} used a very small aperture centered on the peak of the knot to measure the polarization; within such a small region the Doppler depolarization effect is not significant, especially when the flow is only mildly relativistic. But the high polarization can still be challenging in terms of particle acceleration models that require turbulence at the shock. Undoubtedly, the inner knot of the Crab Nebula will be observed more and the conclusions on which we have based our analysis will be corroborated or modified.

Despite the somewhat inconclusive nature of our investigation, it is apparent that observations of AGN jets, $\gamma$-ray bursts and pulsar wind nebulae, especially the Crab Nebula, is rapidly enhancing our appreciation and understanding of how relativistic plasma behave in cosmic sources. Interestingly, similar knot structures are also seen in Chandra images of Vela \citep{Levenfish:2013aa}, the only other pulsar wind nebula which we see at comparable resolution. This may indicate the generality of some features and could give further hints. We believe future multiwavelength monitoring campaign of the Crab and other related sources will help us understand many of the aforementioned open questions.

\section*{Acknowledgements}

We thank Jon Arons, Rolf Buehler, Stefan Funk, Jeff Kolodziejczak, Serguei Komissarov, Maxim Lyutikov, Claire Max, Stephen O'Dell, Oliver Porth, Roger Romani, Alexander Rudy, Jeff Scargle, and Martin Weisskopf for helpful discussions. This work was supported in part by the U.S. Department of Energy contract to SLAC no. DE-AC02-76SF00515, NSF grant AST 12-12195, as well as the Simons Foundation, the Humboldt Foundation, and the Miller Foundation (RB). YY gratefully acknowledges support from the KIPAC Gregory and Mary Chabolla fellowship and the Gabilan Fellowship awarded by Stanford University.




\bibliographystyle{mn2e}
\bibliography{CrabRef} 



\appendix

\section{Generalized shock jump condition}\label{sec:generalized_shock}
\subsection{Anisotropic plasma pressure}\label{subsec:aniso_shock}
\begin{figure}
  \centering
  \includegraphics[width=\columnwidth]{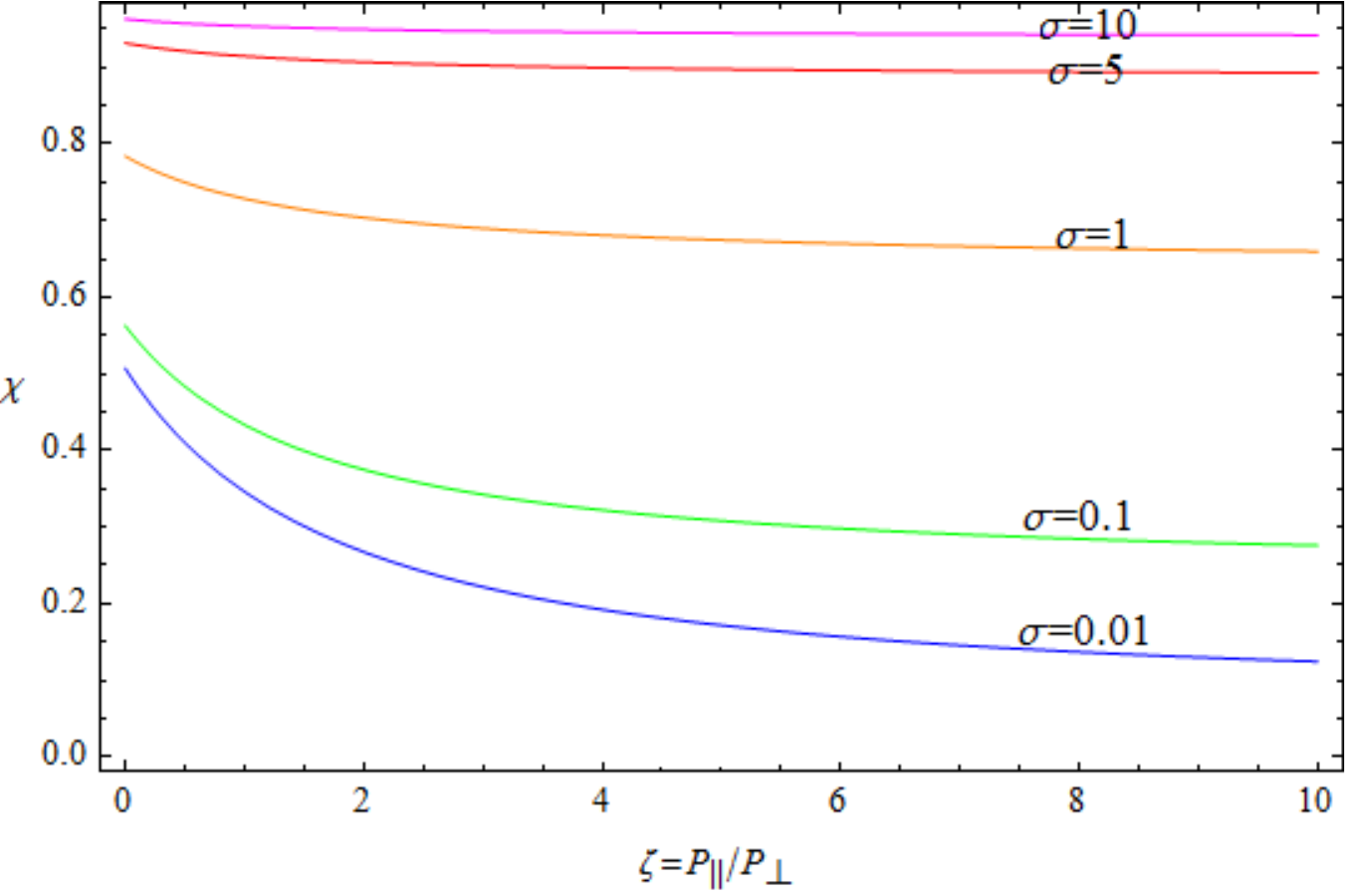}\\
  \caption{Shock compression ratio $\chi$ as a function of the anisotropy parameter $\zeta=P_{\parallel}/P_{\perp}$, for different $\sigma$'s.}\label{fig:aniso_shock}
\end{figure}

In the general case when the particle distribution is gyrotropic but anisotropic between the directions perpendicular and parallel to local magnetic field, the energy-stress tensor for the fluid part can be written as $T_{\text{MA}}^{\mu \nu }=\left(\rho +P_{\perp}\right)u^{\mu }u^{\nu }+P_{\perp}g^{\mu \nu }+\left(P_{\parallel}-P_{\perp}\right)b^{\mu }b^{\nu }$, where $b^{\mu}$ is a unit 4-vector along the direction of magnetic field in the fluid rest frame. The energy density is $\rho =n m c^2+2P_{\perp}+P_{\parallel}$. First consider a perpendicular shock where the incoming velocity is along the shock normal and magnetic field is perpendicular to the shock normal. Applying conservation laws at the shock front gives
\begin{gather}
n_1u_1=n_2u_2,\\
v_1B_1=v_2B_2,\\
\left(\rho _1+P_{1\perp}\right)\gamma _1 u_1+\frac{1}{4\pi }v_1B_1^2=\left(\rho _2+P_{2\perp}\right)\gamma _2 u_2+\frac{1}{4\pi }v_1B_1^2,
\end{gather}
\begin{multline}
\left(\rho _1+P_{1\perp}\right) u_1^2+P_{1\perp}+\frac{1}{8\pi }\left(v_1^2+1\right)B_1^2=\\\left(\rho _2+P_{2\perp}\right) u_2^2+P_{2\perp}+\frac{1}{8\pi }\left(v_2^2+1\right)B_2^2.
\end{multline}
Suppose that the upstream flow is cold ($P_{1\perp}=P_{1\parallel}=0$) and ultrarelativistic ($u_1\approx \gamma _1$, $v_1\approx1$), and define $\sigma =B_1^2/(4\pi  n_1\gamma _1^2m c^2v_1)$. Let $P_{2\parallel}=\zeta  P_{2\perp}$, then $\rho _2=n_2 m c^2+(2+\zeta )P_{2\perp}\approx (2+\zeta )P_{2\perp}$ under strong shock assumptions. Now the above system of equations can be solved to get the compression ratio
\begin{equation}
\begin{split}
\chi& \approx v_2\\
&=\frac{2+3 \sigma +\zeta  \sigma +\sqrt{(2+5\sigma+3\zeta\sigma)^2+8\sigma(1+\zeta)^2}}{4 (2+\zeta)(1+  \sigma )}
\end{split}
\end{equation}
When $\zeta=1$, we get back to the familiar result of isotropic case, Equation (\ref{eq:chi}). In the limit $\zeta=0$ (no parallel pressure), we get $\chi =(2+3 \sigma +\sqrt{4+28 \sigma +25 \sigma ^2})/(8+8 \sigma )$. $\chi \to 1/2+3 \sigma /4$ when $\sigma\to0$ and $\chi\to1-2/(5\sigma)$ when $\sigma\to\infty$. In another limit $\zeta\to\infty$ (no perpendicular pressure) we have $\chi =(\sigma +\sqrt{\sigma  (8+9 \sigma )})/(4 +4\sigma )$. $\chi \to \sqrt{\sigma }/\sqrt{2}+\sigma /4$ as $\sigma\to 0$ and $\chi\to1-2/(3\sigma)$ as $\sigma\to\infty$. Figure \ref{fig:aniso_shock} shows how the compression ratio $\chi$ changes with $\zeta$, for different $\sigma$.
If the upstream plasma is warm and anisotropic, we just need to change the definition of $\sigma$ into $\sigma=B_1^2/(4\pi(\rho_1+P_{1\perp})\gamma_1^2)$, where $\rho_1+P_{1\perp}\equiv w_1$ is the upstream enthalpy density.

When the incoming velocity is oblique with respect to the shock front, it is easily shown that the component of velocity parallel to the shock surface is conserved so one can make a Lorentz transformation to get back to the perpendicular case. The relations for deflection angle and downstream Lorentz factor, Equations (\ref{eq:deflection_angle}) and (\ref{eq:gamma2}), remain valid.

\subsection{The upstream magnetic field has both a $\phi$ and a $\theta$ component, and is possibly striped}\label{subsec:shock_striped}
\begin{figure}
  \centering
  \includegraphics[width=\columnwidth]{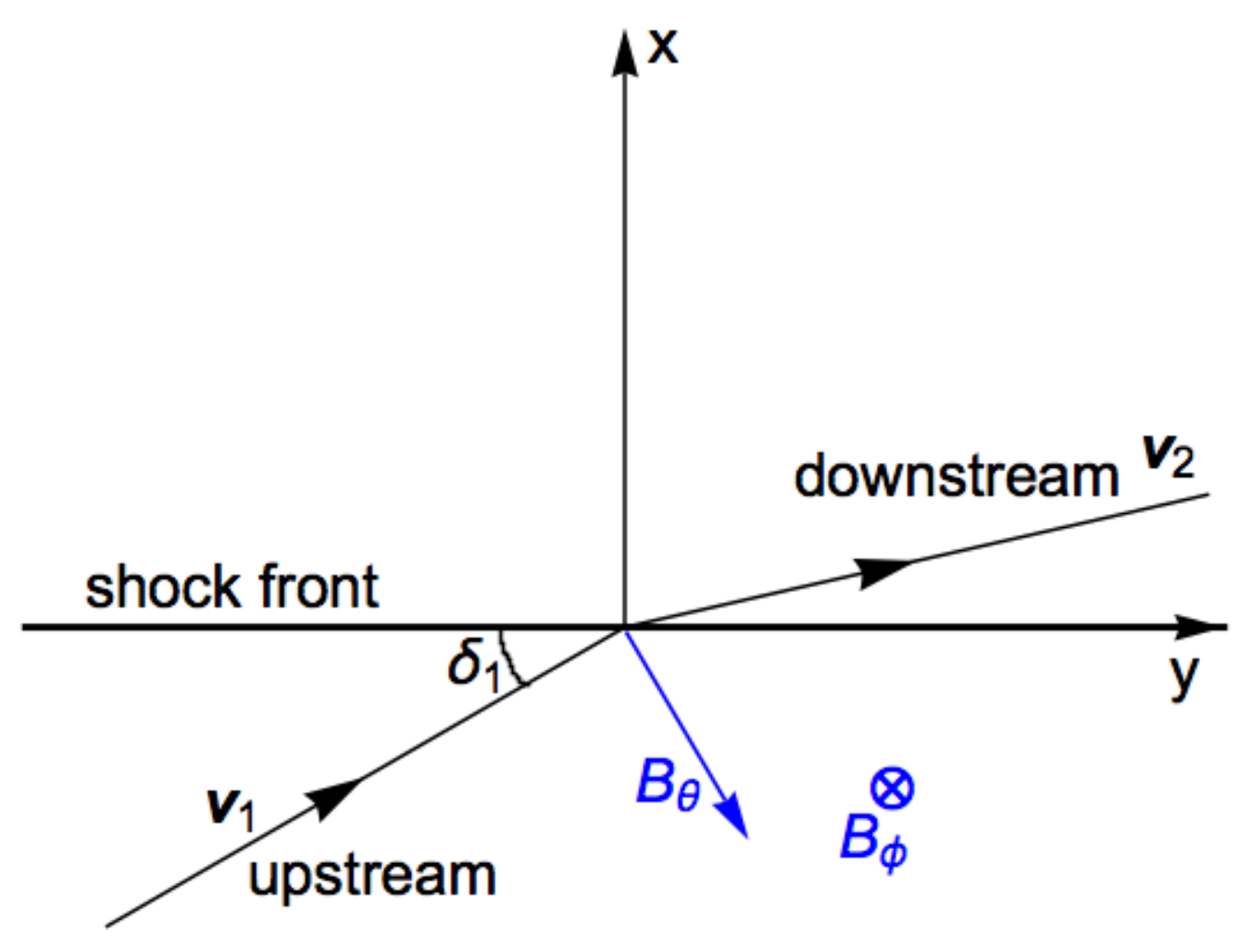}\\
  \caption{Local coordinate system for deriving the shock jump condition in a striped wind.}\label{fig:shock_stripedwind}
\end{figure}

We use a local coordinate system at the shock front, as shown in Figure \ref{fig:shock_stripedwind}, where $\hat{x}$ corresponds to the shock normal, $\hat{z}$ corresponds to $\phi$ direction and $\mathbf{v}_1$ is in radial direction. We suppose that the magnetic field upstream has both a $\theta$ component and a $\phi$ component, while its r component is zero. Consider isotropic plasma first: $T_{\text{MA}}^{\mu \nu }=P g^{\mu \nu }+(P+\rho ) u^{\mu } u^{\nu }$, and let the adiabatic index to be $\Gamma=4/3$. For the upstream stress tensor, we average over a stripe wavelength. Continuity of the stress tensor components $T^{\mu  x}$, field components $B_x$, $E_y$, $E_z$ and particle flux across the shock front give the following equations:
\begin{equation}
\begin{split}
T^{0x}:\ &\left\langle \frac{B_{1 y} \left(B_{1 y} v_{1 x}-B_{1 x} v_{1 y}\right)+B_{1 z}^2 v_{1 x}}{4 \pi }+\gamma _1 \left(P_1+\rho _1\right) u_{1 x}\right\rangle\\
 &=\frac{B_{2 y} \left(B_{2 y} v_{2 x}-B_{2 x} v_{2 y}\right)-B_{2 z} \left(B_{2 x} v_{2 z}-B_{2 z} v_{2 x}\right)}{4 \pi }\\
 &+\gamma _2 \left(P_2+\rho _2\right) u_{2 x},
\end{split}
\end{equation}
\begin{equation}
\begin{split}
T^{xx}:\ &\left\langle -\frac{1}{4\pi} \left(B_{1 z}^2 v_{1 y}^2+B_{1 x}^2\right)+\frac{1}{8\pi}\left(B_{1 y} v_{1 x}-B_{1 x} v_{1 y}\right)^2\right\rangle\\
&+\left\langle\frac{1}{8 \pi }(B_{1 z}^2 v_{1 x}^2+B_{1 z}^2 v_{1 y}^2+B_1^2)+\left(P_1+\rho _1\right) u_{1 x}^2+P_1\right\rangle\\
 &=\left(P_2+\rho _2\right) u_{2 x}^2+P_2\\
 &+\frac{1}{8 \pi }\left[-2 \left(\left(B_{2 z} v_{2 y}-B_{2 y} v_{2 z}\right)^2+B_{2 x}^2\right)+\left(B_{2 y} v_{2 x}-B_{2 x} v_{2 y}\right)^2\right.\\
 &\left.+\left(B_{2 x} v_{2 z}-B_{2 z} v_{2 x}\right)^2+\left(B_{2 z} v_{2 y}-B_{2 y} v_{2 z}\right)^2+B_2^2\right],
 \end{split}
\end{equation}
 \begin{equation}
\begin{split}
T^{xy}:\ &\left\langle \left(P_1+\rho _1\right) u_{1 x} u_{1 y}-\frac{B_{1 x} B_{1 y}-B_{1 z}^2 v_{1 x} v_{1 y}}{4 \pi }\right\rangle\\
 &=-\frac{\left(B_{2 x} v_{2 z}-B_{2 z} v_{2 x}\right) \left(B_{2 z} v_{2 y}-B_{2 y} v_{2 z}\right)+B_{2 x} B_{2 y}}{4 \pi }\\
 &+\left(P_2+\rho _2\right) u_{2 x} u_{2 y},
\end{split}
\end{equation}
\begin{equation}
\begin{split}
T^{xz}:\ &\left\langle -\frac{v_{1 y} B_{1 z} \left(B_{1 y} v_{1 x}-B_{1 x} v_{1 y}\right)+B_{1 x} B_{1 z}}{4 \pi }\right\rangle\\
 &=-\frac{\left(B_{2 y} v_{2 x}-B_{2 x} v_{2 y}\right) \left(B_{2 z} v_{2 y}-B_{2 y} v_{2 z}\right)+B_{2 x} B_{2 z}}{4 \pi }\\
 &+\left(P_2+\rho _2\right) u_{2 x} u_{2 z},
\end{split}
\end{equation}
\begin{gather}
\left\langle B_{1 x}\right\rangle =B_{2 x},\\
\left\langle E_{1 y}\right\rangle =E_{2 y}
\text{, or}
\left\langle B_{1 z} v_{1 x}\right\rangle =B_{2 z} v_{2 x}-B_{2 x} v_{2 z},\\
\left\langle E_{1 z}\right\rangle =E_{2 z}
\text{, or}
\left\langle B_{1 y} v_{1 x}-B_{1 x} v_{1 y}\right\rangle =B_{2 y} v_{2 x}-B_{2 x} v_{2 y},\\
n_1 u_{1 x}=n_2 u_{2 x}
\end{gather}
Here $\langle \rangle$ denotes an average over a wavelength. Suppose that the upstream flow is ultra-relativistic, namely $v_1\approx 1$, $\gamma _1\approx u_1\gg 1$. We can write $v_{1 x}=\sin\delta _1 $, $v_{1 y}=\cos\delta _1 $. From the geometry as shown in Figure \ref{fig:shock_stripedwind}, we have $B_{1 z}=B_{\phi }$, $B_{1 x}= -B_{\theta }\cos \delta _1$, $B_{1 y}=B_{\theta }\sin  \delta _1 $. The solution to the above equations satisfies $v_{2z}=0$, and the compression ratio is
\begin{equation}
\chi \equiv\frac{v_{2x}}{v_{1x}}=\frac{\sqrt{\sigma ^2 \varsigma^2+14 \sigma ^2 \varsigma+\sigma ^2+14 \sigma  \varsigma+2 \sigma +1}+\sigma  \varsigma+\sigma +1}{6 (\sigma +1)}
\end{equation}
where $\sigma$ is defined as
\begin{equation}
\sigma \equiv\frac{\left\langle B_{\theta }^2\right\rangle +\left\langle B_{\phi }^2\right\rangle }{4 \pi \gamma _1^2 \left(P_1+\rho _1\right)}
\end{equation}
and $\varsigma\equiv\left(\left\langle B_{\theta }\right\rangle^2+\left\langle B_{\phi }\right\rangle {}^2\right)/\left(\left\langle B_{\theta }^2\right\rangle +\left\langle B_{\phi }^2\right\rangle \right)$. If we let $\tilde{\sigma }=\sigma  \varsigma/(\sigma +1)$, we get
\begin{equation}
\chi =\frac{1}{6} \left(1+\tilde{\sigma }+\sqrt{\tilde{\sigma }^2+14 \tilde{\sigma }+1}\right),
\end{equation}
generalizing the results obtained by \citet{Lyubarsky:2003aa}.


\bsp	
\label{lastpage}
\end{document}